\documentstyle[aps,graphicx,epsf]{revtex}\tighten
\begin{document}
\def\tablerule{\noalign {\vskip3truept\hrule\vskip3truept}}
\def\half{{1\over 2}}
\def \e {\eta }
\def \ee {({\eta\over \eta _0})}
\def \D {\mbox{D}}
\def\curl {\mbox{curl}\,}
\def \ep {\varepsilon}
\def \lleq {\lower0.9ex\hbox{ $\buildrel < \over \sim$} ~}
\def \ggeq {\lower0.9ex\hbox{ $\buildrel > \over \sim$} ~}
\newcommand{\sq}{\lower.25ex\hbox{\large$\Box$}}
\def \l {\Lambda}
\def\beq{\begin{equation}}
\def\eeq{\end{equation}}
\def\ber{\begin{eqnarray}}
\def\eer{\end{eqnarray}}
\def \ie {{\em i.e.~~}}
\def \apl {ApJ, }
\def \aps {ApJS, }
\def \pd {Phys. Rev. D, }
\def \prl {Phys. Rev. Lett., }
\def \pl {Phys. Lett., }
\def \np {Nucl. Phys., }

\draft

\title{Relic Gravity Waves from Braneworld Inflation}

\author{Varun Sahni$^{1,*}$, M. Sami$^{2,\dagger}$ and Tarun
Souradeep$^{1,\ddagger}$}
\address{$^1$Inter-University Centre for
Astronomy and Astrophysics, Ganeshkhind, Pune 411 007, India}
\address{$^2$ Department of Physics, Jamia Millia Islamia, New Delhi 110025, India}

\maketitle

\begin{abstract}
We discuss a scenario in which extra dimensional
effects allow a scalar field with a steep potential
to play the dual role of the inflaton as well as dark energy (quintessence).
The post-inflationary evolution of the universe in this scenario is
generically characterised by a `kinetic regime' during which the kinetic energy
of the scalar field greatly exceeds its potential energy
resulting in a
`stiff' equation of state for scalar field matter $P_\phi \simeq \rho_\phi$.
The kinetic regime precedes the radiation dominated epoch and
introduces an important new feature into the spectrum of relic gravity waves
created quantum mechanically during inflation. The amplitude of the 
gravity wave
spectrum {\em increases with wavenumber} for wavelengths shorter than the
comoving horizon scale at the commencement of the radiative regime.
This `blue tilt' is a generic feature of
models with steep potentials and imposes strong constraints on a
class of inflationary braneworld models.
Prospects for detection of the gravity wave background by terrestrial and
space-borne gravity wave
observatories such as LIGO II and LISA are discussed.
\end{abstract}


\section{Introduction}
The last two decades have seen considerable effort being devoted to the
construction of fundamental theories of nature in more than three
spatial dimensions. In such models the four
dimensional Planck scale $M_4 \equiv G^{-1/2}
= 1.2 \times 10^{19}$ GeV is related to its fundamental value $M_{\rm f}$
by $M_4^2 = M_{\rm f}^{2+n} {\cal R}^n$ where $n$ is the number of
extra dimensions.
In the original Kaluza-Klein picture the
extra dimensions were compact and microscopic $R \sim 10^{-33}$ cm,
hence unobservable.
However  it was soon realized that a theory in which one or more of the extra
dimensions is
macroscopic ($R \sim 1$ mm) has several interesting features. For instance
a model in which two of the extra dimensions extend up to a  millimeter
has the considerable
advantage of decreasing the fundamental Planck scale to
electroweak scales
$M_{\rm f} \sim 1 ~{\rm TeV} \ll M_4$ thereby alleviating the hierarchy
problem associated with particle physics \cite{dvali}.
A further
development of these ideas \cite{rubakov83,rs} led to a scenario in which
our universe is a three dimensional domain wall (brane) embedded
in an infinite four dimensional space (bulk).
The metric describing
the full 4+1 dimensional space-time is non-factorizable and the small
value of the true five dimensional Planck mass is related to its large
effective four dimensional value by the extremely large warp of five
dimensional space \cite{rs}.
Since gravity and matter fields remain confined to
the brane the presence of the extra
(bulk) dimension does not affect Newton's law which remains
inverse square.

As demonstrated in \cite{brane,maartens} the prospects of inflation
in such a scenario improve due to the presence of an additional
quadratic density term in the Einstein equations.
The class of potentials which lead to inflation increases and
includes potentials which are normally too steep to be associated with
inflation \cite{copeland}.
As we demonstrate in this paper, for a suitable choice of parameter
values, scalar field with steep potentials can successfully play the
dual role of being the inflaton at early times and dark energy (quintessence)
at late times (see also \cite{pow01}). 
In inflationary models with steep potentials the immediate aftermath of
braneworld inflation is characterised by a  `kinetic regime' during which the inflaton
has the `stiff' equation of state $p \simeq \rho$ \cite{zel}.

In this paper we examine an important observational imprint of
brane-inflation - the spectrum of relic gravity waves.
The quantum mechanical creation of gravitons from the vacuum is an
important generic feature of expanding cosmological models
\cite{grish}.
In models characterised by an early inflationary epoch
the relic gravity wave amplitude is related to the Hubble
parameter during inflation \cite{star} while its spectrum
depends upon both the inflationary and post-inflationary equation of state
\cite{star,allen,sahni}. 
As demonstrated in \cite{lmw00} the gravity wave amplitude
in braneworld models is considerably enhanced over that in standard
inflation. Additionally,
as we show in this paper, the presence
of a kinetic regime soon after brane-inflation
leads to a `blue' spectrum for
gravity waves, on scales smaller than 
the comoving horizon scale at the commencement of the radiative regime
(see also \cite{giovan}).
This unique feature of steep braneworld inflation results in an
enhanced gravity wave background which can be used to succesfully
constrain braneworld models. 

\section{Braneworld Inflation}
In the 4+1 dimensional brane scenario
inspired by the Randall-Sundrum \cite{rs} model, the standard 0-0 Friedman
equation is modified to \cite{brane}
\beq
H^2 = \frac{8\pi}{3 M_4^2}\rho (1 + \frac{\rho}{2\lambda_b}) + \frac{\Lambda_4}{3} +\frac{\cal E}{a^4}
\label{eq:frw1}
\eeq
where ${\cal E}$ is an integration constant which transmits bulk graviton
influence onto the brane and
$\lambda_b$ is the three dimensional brane tension which provides a
relationship between the four and five-dimensional Planck masses
\beq
M_4 = \sqrt{\frac{3}{4\pi}}\big (\frac{M_5^2}{\sqrt{\lambda_b}}\big )M_5,
\eeq
and also relates the four-dimensional cosmological constant $\Lambda_4$
to its five-dimensional counterpart via
\beq
\Lambda_4 = \frac{4\pi}{M_5^3}\bigg (\Lambda_5 + \frac{4\pi}{3M_5^3}\lambda_b^2
\bigg).
\eeq
Following \cite{maartens}
we make the additional assumption that $\Lambda_4$ is too small
to play an important role in the early universe (the possible presence of
a small cosmological constant today
provides the lower bound
$\Omega_\Lambda = M_4^2 \Lambda_4/8\pi \sim 0.7$). The ``dark radiation''
term ${\cal E}/a^4$ is expected to rapidly disappear once inflation has
commenced so that we effectively get \cite{brane,maartens}
\beq
H^2 = \frac{8\pi}{3 M_4^2}\rho (1 + \frac{\rho}{2\lambda_b}),
\label{eq:frw2}
\eeq
where $\rho = \half{\dot\phi}^2 + V(\phi)$, if one is dealing with a
universe dominated by a single minimally coupled scalar field.
The equation of motion of a scalar field propogating on the brane is
\begin{equation}
{\ddot \phi} + 3H {\dot \phi} + V'(\phi) = 0.
\label{eq:kg}
\end{equation}
From (\ref{eq:frw2}) and (\ref{eq:kg}) we find that the presence of the additional term
$\rho^2/\lambda_b$
increases the damping experienced by the scalar field as it rolls down its
potential. This effect is reflected in the slow-roll parameters which
have the form \cite{maartens}
\ber
\epsilon &=& \epsilon_{\rm FRW}
\frac{1 + V/\lambda_b}{(1 + V/2\lambda_b)^2},\nonumber\\
\eta &=& \eta_{\rm FRW}(1 + V/2\lambda_b)^{-1},
\label{eq:slow-rollbrane}
\eer
where
\beq
\epsilon_{\rm FRW} = \frac{M_4^2}{16\pi} \left (\frac{V'}{V}\right )^2,
\,\, \eta_{\rm FRW} = \frac{M_4^2}{8\pi} \left (\frac{V''}{V}\right )
\label{eq:slow_FRW}
\eeq
are slow roll parameters in the absence of brane corrections.
The influence of the brane term becomes important when $V/\lambda_b \gg 1$
and in this case we get
\beq
\epsilon \simeq 4\epsilon_{\rm FRW} (V/\lambda_b)^{-1},\,
\eta \simeq 2\eta_{\rm FRW} (V/\lambda_b)^{-1}.
\label{eq:slow-roll}
\eeq
Clearly slow-roll ($\epsilon, \eta \ll 1$) is easier to achieve when
$V/\lambda_b \gg 1$ and on this basis one can expect inflation
to occur even for relatively steep
potentials, such the exponential and the inverse power-law which we discuss
below.

\subsection{Exponential potentials}
\label{sec:exp}

The exponential potential
\beq
V(\phi) = V_0e^{\tilde\alpha\phi/M_P}
\label{eq:exponent}
\eeq
with ${\dot \phi} < 0$
(equivalently $V(\phi) = V_0e^{-\tilde\alpha\phi/M_P}$
with ${\dot \phi} > 0$)
has traditionally played an important role
within the inflationary framework \cite{lm85} since, in the absence of matter,
it gives rise to power law inflation $a \propto t^c$,
$c = 2/\tilde\alpha^2$
provided $\tilde\alpha\leq \sqrt{2}$.
($M_P$ is the reduced four dimensional Planck mass $M_P = M_4/\sqrt{8\pi}$,
$M_4 = 1.2\times 10^{19}$ GeV.)
For $\tilde\alpha > \sqrt{2}$
the potential becomes too steep to sustain inflation and for
larger values $\tilde\alpha \geq \sqrt{6}$ the field enters a
kinetic regime during which
${\dot\phi}^2 \gg V(\phi)$ and $p_\phi \simeq \rho_\phi$,
resulting in a rapidly
decreasing field energy density $\rho_\phi \propto a^{-6}$.
Thus within the standard general relativistic framework,
steep potentials which have been suggested as candidates for cold dark matter and
quintessence \cite{rp,exp1,exp2,sw99}
are not capable of sustaining inflation.
However
extra-dimensional effects lead to interesting new possibilities for the
inflationary scenario. The increased damping of the scalar field
when $V/\lambda_b \gg 1$ leads to
a decrease in the value of the slow-roll parameters
$\epsilon = \eta \simeq 2\tilde\alpha^2\lambda_b/V$,
so that slow-roll ($\epsilon,\eta \ll 1$) leading to inflation
now becomes possible even for large values of $\tilde\alpha$.

Within the framework of the braneworld scenario, the field equations
(\ref{eq:frw2}) and (\ref{eq:kg}) can be solved {\em exactly} in the
slow-roll limit when
$\rho/\lambda_b \gg 1$. In this case
\beq
\frac{\dot{a}(t)}{a(t)} \simeq \frac{1}{\sqrt{6 M_P^2\lambda_b}} V(\phi),
\label{eq:frw3}
\eeq
which, when substituted in
\beq
3H\dot{\phi} \simeq -V'(\phi)
\label{eq:kg1}
\eeq
leads to ${\dot\phi} = - \tilde\alpha\sqrt{2\lambda_b/3}$.
The equation of motion for the $\phi$-field follows immediately
\beq
\phi(t)= \phi_i - \sqrt{\frac{2\lambda_b}{3}} \tilde\alpha (t-t_i).
\label{eq:motion}
\eeq
An expression for number of inflationary e-foldings is also easy to establish
\ber
{\cal N} &=& \log{\frac{a(t)}{a_i}} = \int_{t_i}^t H(t') dt'
\simeq \frac{V_0}{2\lambda_b\tilde\alpha^2}(e^{\tilde\alpha\phi_i} - e^{\tilde\alpha\phi(t)})\label{eq:efold0}\\
&=& \frac{V_i}{2\lambda_b\tilde\alpha^2}\left [ 1 -
\exp{\lbrace-\sqrt{\frac{2\lambda_b}{3M_P^2}\tilde\alpha^2}(t-t_i)\rbrace}\right ],
\label{eq:efold}
\eer
where $V_i = V_0e^{\tilde\alpha\phi_i}$.
From Eq.~(\ref{eq:efold}) we find that the expansion factor passes
through an inflection point at $$t-t_i = \gamma =
\sqrt{\frac{3M_P^2}{2\lambda_b}}\tilde\alpha^{-2}
\log{\bigg (\frac{V_i}{2\lambda_b\tilde\alpha^2}\bigg )},$$
since
\ber
{\ddot a} &>& 0 ~~{\rm for}~~ t-t_i < \gamma,\nonumber\\
{\ddot a} &<& 0 ~~{\rm for}~~ t-t_i > \gamma, \nonumber\\
{\ddot a} &=& 0 ~~{\rm for}~~ t-t_i = \gamma.
\eer
The epoch
$t_{\rm end} = t_i + \gamma$ therefore marks the end of inflation.
Substitution in (\ref{eq:motion}) and (\ref{eq:exponent}) gives
\beq
\phi_{\rm end} =
-\frac{M_P}{\tilde\alpha}
\log{\bigg (\frac{V_0}{2\lambda_b\tilde\alpha^2}\bigg )},
\label{eq:phiend}
\eeq
\beq
V_{\rm end} \equiv V_0e^{\tilde\alpha\phi_{\rm end}/M_P} =
2\lambda_b\tilde\alpha^2\\
\label{eq:Vend}.
\eeq
From  (\ref{eq:efold0}) \& (\ref{eq:Vend}) we also find
\ber
{\cal N} = \log{a/a_i} &=& V_0\big \lbrack e^{\tilde\alpha\phi_i/M_P}
- e^{\tilde\alpha\phi(t)/M_P}\big \rbrack/V_{\rm end}\label{eq:efold1},\\
{\rm or}~~  V_{\rm end} = \frac{V_i}{{\cal N}+1}.
\label{eq:efold2}
\eer
Eqns.~(\ref{eq:Vend}) \& (\ref{eq:efold2})
had earlier been obtained in \cite{copeland} using a different method.
The COBE normalized value for the amplitude of scalar density perturbations
\beq
A_s^2 \simeq \frac{8}{75}\frac{V_i^4}{M_4^4\tilde\alpha^2\lambda_b^3} \simeq 4 \times 10^{-10}
\label{eq:cobe}
\eeq
can now be used to determine both
$\lambda_b$ and $V_{\rm end}$:
\ber
\lambda_b &\simeq& \frac{2.3\times 10^{-10}}{\tilde\alpha^6}
\left (\frac{M_4}{{\cal N}+1}\right )^4,\\
V_{\rm end} &\simeq&
4.6\times 10^{-10} \left (\frac{M_4}{\tilde\alpha({\cal N}+1)}\right )^4,
\label{eq:tension}
\eer
in agreement with the results obtained in \cite{copeland}.
For ${\cal N} \simeq 70$ we get
\ber
\lambda_b &\simeq& 9.3\times 10^{-18}\frac{M_4^4}{\tilde\alpha^6}
\simeq \frac{1.9\times 10^{59}}{\tilde\alpha^6} ~{\rm GeV}^4,\\
V^{1/4}_{\rm end} &\simeq& \frac{6.5\times 10^{-5}}{\tilde\alpha} M_4 \simeq
\frac{7.8 \times 10^{14}}{\tilde\alpha} ~{\rm GeV}.
\label{eq:tension1}
\eer
(The tensor/scalar ratio in the CMB anisotropy in braneworld
inflation is small \cite{lmw00} and we work under the assumption that scalar density
perturbations are responsible for most of the COBE signal.)

From (\ref{eq:Vend}) we find that
\beq
\rho_{\rm end}/2\lambda_b \simeq V_{\rm end}/2\lambda_b = \tilde\alpha^2, 
\label{eq:tension2}
\eeq
and
\ber
H_{\rm end} &\simeq& \sqrt{\frac{8\pi V_{\rm end}}{3}}\frac{\tilde\alpha}{M_4},\\
H_i &\simeq& ({\cal N} + 1)H_{\rm end}.
\label{eq:tension2a}
\eer
Clearly $\rho_{\rm end}/2\lambda_b \gg 1$ if ${\tilde\alpha}^2 \gg 1$
which suggests that the brane
term in (\ref{eq:frw2}) will continue to dominate the dynamics of the
universe for some time after inflation has ended, as demonstrated in Fig. ~1.

Eqs.~(\ref{eq:efold0}) \& (\ref{eq:efold}) can be rewritten as
\ber
{\cal N} &=& \log{\frac{a(t)}{a_i}} = \frac{V_i}{V_{\rm end}}\left [
1 - \exp{\frac{\tilde\alpha}{M_P}\left \{\phi(t) - \phi_i)\right \}}\right ] \\
&=& \frac{V_i}{V_{\rm end}}\left [ 1 -
\exp{\left \{- \sqrt{\frac{2\lambda_b}{3M_P^2}}\tilde\alpha^2 (t-t_i)\right \}}\right ].
\label{eq:efold3}
\eer
Expanding the right hand side of (\ref{eq:efold3}) in a Taylor series we
get
\ber
\frac{a(t)}{a_i} &\simeq&
\exp{\left [ \frac{V_i}{V_{\rm end}}\frac{\tilde\alpha}{M_P}(\phi_i-\phi(t))\right ]} \\
&=&
\exp{\left [\sqrt{\frac{2\lambda_b}{3M_P^2}}\tilde\alpha^2\frac{V_i}{V_{\rm end}}(t-t_{\rm end})
\right ]}
\label{eq:efold4}
\eer
which demonstrates
that inflation proceeds at an exponential rate during early epochs.


\subsubsection{Post-inflationary evolution: The kinetic regime}

A short while after inflation ends, the brane-term in the field
equations (\ref{eq:frw2}) becomes unimportant. The scalar field rolling down a
steep potential is now subject to minimum damping and soon
goes into a `free fall' mode during which ${\dot\phi}^2 \gg V(\phi)$
and $\rho_\phi \propto a^{-6}$.
Integrating the system of equations (\ref{eq:frw2}) \& (\ref{eq:kg}) 
numerically,
we find that the value of the Hubble parameter at the
commencement of the kinetic regime can be conveniently related to its value 
at the end of inflation by
the fitting formula 
\beq
\frac{H_{\rm kin}}{H_{\rm end}} = a + \frac{b}{\tilde\alpha^2}, ~~
(\tilde\alpha \ggeq 3)
\label{eq:fit}
\eeq
where $a = 0.085$ and $b = -0.688$.
In addition, a small amount of
radiation is also present due to particles being produced quantum mechanically
during inflation which give rise to an energy density\cite{ford,spokoiny}
$\rho_R \sim 0.01 g_p H_{\rm end}^4$,
where $g_p \simeq 10 - 100$ is the number of different particle species
created from the vacuum.
(In the case of the exponential potential, quantum mechanical particle
production provides the only mechanism by means of which the universe
can `reheat'.)
If the particles created quantum mechanically
were to be thermalized immediately after inflation, then one would obtain
for the radiation temperature the value \cite{copeland,reheat}
\beq
T_{\rm end} \simeq \frac{2 \times 10^{-5}}{\tilde\alpha}\frac{M_4}
{({\cal N} + 1)^2} = \frac{9 \times 10^{10}}{\tilde\alpha}\left (
\frac{51}{{\cal N} + 1}\right )^2 {\rm GeV}.
\label{eq:temp0}
\eeq
It can be easily shown that in this case
the density in the inflaton far exceeds the density in radiation at the
end of inflation
\beq
\left (\frac{\rho_\phi}{\rho_r}\right)_{\rm end} \simeq 5\lambda_b^2M_4^4/g_pV_{\rm end}^3 \sim 2 \times 10^{16}\left (\frac{{\cal N} + 1}{51}\right )^4
g_p^{-1}.
\label{eq:ratio_end}
\eeq
This leads to a prolonged `kinetic regime' during which scalar matter has
the `stiff' equation of state $P_\phi
\simeq \rho_\phi$ and the universe evolves as
\beq
\frac{a(t)}{a_{\rm kin}}=
\bigg\lbrack\sqrt{\frac{\tilde\alpha^2\lambda_b}{M_P^2}}(t-t_{\rm kin})
+ 1\bigg\rbrack^{1/3},
\label{eq:evol_kin}
\eeq
while the evolution of the scalar field is described by
\beq
\phi(t)= \phi_{\rm kin} -\sqrt 6 M_P \log{\left({a(t) \over
a_{\rm kin}}\right)}.
\eeq
Here $a_{\rm kin}, \phi_{\rm kin}$ correspond to
values at the start of the kinetic regime and one can
assume to a first approximation
$\phi_{\rm kin} \simeq \phi_{\rm end}$,
$a_{\rm kin} \simeq a_{\rm end}$, although, as we demonstrate below, 
the kinetic regime does not commence immediately after inflation ends 
but shortly afterwards.
After the commencement of the kinetic regime,
radiation and inflaton densities fall off at different rates
$\rho_{\rm rad}/\rho_\phi \propto a^2$ and a time will come when
the two will equalize
\beq
\left (\frac{\rho_\phi}{\rho_r}\right )_{\rm eq}
 = \left (\frac{\rho_\phi}{\rho_r}\right )_{\rm kin}
({a_{\rm kin} \over a_{\rm eq}})^2 \simeq 1.
\eeq
The value of the scalar field at this juncture is given by
\beq
{\phi_{eq} \over M_P}= \frac{\phi_{\rm kin}}{M_P}
-{\sqrt6 \over 2}\log{\left (\frac{\rho_\phi}{\rho_r}\right )_{\rm kin}}
\eeq
and the temperature of the universe can be estimated from
\beq
T_{\rm eq} = T_{\rm kin}\left (\frac{a_{\rm kin}}{a_{\rm eq}}\right )
= T_{\rm kin}\left (\frac{\rho_r}{\rho_\phi}\right )^{1/2}_{\rm kin}
\label{eq:temp}
\eeq
where $T_{\rm kin}$ is the temperature at the start
of the kinetic regime. 
As demonstrated in Fig.~1 the influence of the brane term in (\ref{eq:frw2})
causes a short delay between the commencement of the kinetic regime 
and the end of inflation.

\vfill\eject
\begin{figure}[h]
\centering
\resizebox{!}{2.5in}{\includegraphics{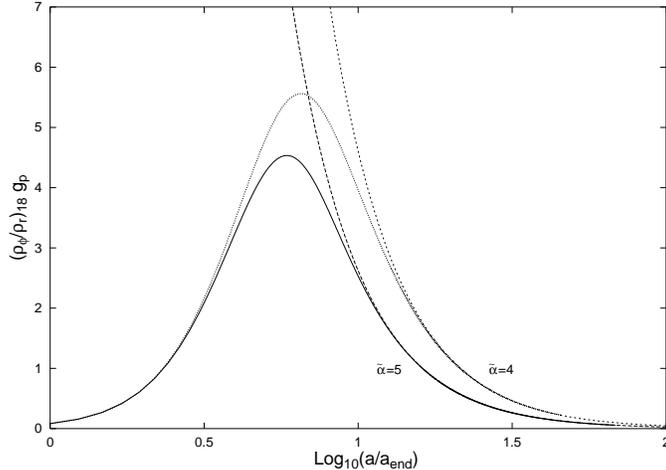}}
\caption{
The evolution of $(\rho_\phi/\rho_{\rm r})_{18} = \rho_\phi/\rho_{\rm r}
\times 10^{-18}$
is shown as a function of the expansion factor shortly after inflation ends.
The ratio $\rho_\phi/\rho_{\rm rad}$ first increases due to the dominance of the
brane-term which causes the density in the $\phi$-field to decrease much more
slowly than the density in radiation.
The decay law $\rho_\phi/\rho_{\rm rad} \propto a^{-2}$ marking the commencement
 of the
kinetic regime is shown for comparison (dashed line). We see that the
influence of the brane term is stronger for {\em smaller}
value of the parameter $\tilde\alpha$.}
\end{figure}

When the kinetic regime finally commences the temperature of the universe has
dropped to
\beq
T_{\rm kin} = T_{\rm end}\left (\frac{a_{\rm end}}{a_{\rm kin}}\right )
\simeq T_{\rm end}\left ( c + \frac{d}{\tilde\alpha^2}\right )
\label{eq:temp1}
\eeq
where $c \simeq 0.142, ~d = -1.057$ and $\tilde\alpha \ggeq 3$ is assumed.
From this expression and Fig. ~1
we see that the influence of the brane term is {\em larger} for
smaller values of $\tilde\alpha$. (For very small values of
$\tilde\alpha$ the temperature at the onset of the kinetic regime $T_{\rm kin}$
can fall
by almost two orders of magnitude relative to $T_{\rm end}$.)
The reason for this can be understood
by rewriting (\ref{eq:frw2}) as
\beq
H^2 = \frac{8\pi}{3 M_4^2}\rho B_{\rm rane}
\eeq
where $B_{\rm rane} = 1 + \rho/2\lambda_b$ is the brane-correction term to the
FRW equations. Immediately after inflation the
influence of the brane term is still strong so that
 $\rho \simeq V_0e^{\tilde\alpha\tilde\phi}
$ where $\tilde\phi = \phi/M_P$. Consequently one can rewrite $B_{\rm rane}$
as $B_{\rm rane}(\tilde\alpha) \simeq
1 + \tilde\alpha^2 e^{-\tilde\alpha (\tilde{\phi}_{\rm end} - \tilde{\phi})}$,
from where it becomes clear that as
$\tilde{\phi}$ rolls to smaller values
($\tilde{\phi} < \tilde{\phi}_{\rm end}$) the influence of the brane term
diminishes for larger values of $\tilde\alpha$. This effect is illustrated
in Fig ~1.
From (\ref{eq:temp1}) one also finds for the commencement of
the kinetic regime, the expression
\beq
\frac{a_{\rm kin}}{a_{\rm end}} = \frac{T_{\rm end}}{T_{\rm kin}}
\simeq (c + \frac{d}{\tilde\alpha^2})^{-1},
\eeq
some values of $a_{\rm kin}/a_{\rm end}$ are given in Table ~1.

\begin{table}
\begin{center}
\caption{Commencement of the kinetic regime.}
\bigskip
\begin{tabular}{cccccccc}
$\tilde\alpha$ & 2.6& 3& 4 & 5 &10&   $\infty$ \\
\tablerule
$a_{\rm kin}/a_{\rm end}$ & 1584.9 & 100 & 14.13 & 11.22 & 7.94&7.05 \\
\tablerule
$\left[\left(\rho_{\phi}/\rho_{r}\right)_{kin}\right]\times{g_p}\times10^{-18}$&
0.009&0.24&1.79&2.23&2.35&2.4\\ 
\end{tabular}
\label{table:omega}
\end{center}
\end{table}

The equality between scalar-field matter and radiation can be
estimated from (\ref{eq:temp}). Integrating the equations of motion
numerically, we find that for $\tilde\alpha \ggeq 3$ the
resulting value of $T_{\rm eq}$ is well described by \ber T_{\rm eq}
&\simeq& \frac{T_{\rm end}}{(\rho_\phi/\rho_r)_{\rm end}^{1/2}}
\left (e + \frac{f}{\tilde\alpha^2}\right )\\
&\simeq& \frac{240}{\tilde\alpha}\sqrt{\frac{g_p}{2}}\left ( e +
  \frac{f}{\tilde\alpha^2}\right ) ~{\rm GeV}
\label{eq:temp2}
\eer
where $e = 0.0265, ~f = -0.176$. For $\tilde\alpha > 15$ we get
\beq
T_{\rm eq} \simeq \frac{6}{\tilde\alpha}\sqrt{\frac{g_p}{2}}~{\rm GeV}.
\label{eq:temp01}
\eeq
From (\ref{eq:temp2}) \& (\ref{eq:temp01}) we find that the temperature at
matter-radiation
equality is sensitive to the value of $\tilde\alpha$: for steep potentials
(large $\tilde\alpha$) $T_{\rm eq}$ is smaller as shown in Fig. ~2.
An upper
limit on $\tilde\alpha$ can be established by requiring that the density
in stiff matter has dropped below the radiation density during
cosmological nucleosynthesis:
$T_{\rm eq} > T_{\rm nucl} \sim {\rm few} {\rm ~MeV}$,
substitution in (\ref{eq:temp01}) leads to the generous upper limit
$\tilde\alpha \lleq 10^4$.

\begin{figure}[h]
\centering
\resizebox{!}{2.5in}{\includegraphics{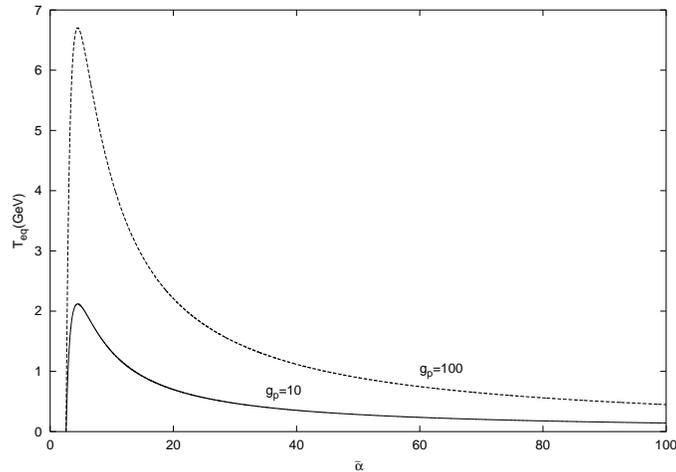}}
\caption{The temperature of the universe at the epoch of radiation
domination (in GeV) is shown as a function of the parameter $\tilde\alpha$
for a universe in which radiation is created due to gravitational
particle production.
}
\end{figure}

The universe has expanded by a factor
$a_{\rm eq}/a_{\rm end} = T_{\rm end}/T_{\rm eq} \sim 10^{10} g_p^{-1/2}$
during the kinetic regime,
between the end of inflation and matter-radiation equality.
As the universe expands further it enters the radiation dominated
regime. As we show in Sec.~(\ref{sec:gravity}) a long kinetic regime
leaves behind a unique signature in the relic gravity wave background
which can be used to strongly constrain inflationary models with
steep potentials.

Finally we should mention that the number of inflationary e-foldings
in this model
can be unambiguously determined from the following considerations.
A length scale which crosses the Hubble radius during the inflationary
epoch ($a_i$) and re-enters it today ($a_0$) will satisfy
$k = a_iH_i = a_0 H_0$, equivalently
\beq
\frac{k}{a_0 H_0} = \frac{a_i}{a_{\rm end}}\frac{a_{\rm end}}{a_0}
\frac{H_i}{H_0} \equiv e^{-\cal N}\frac{T_0}{T_{\rm end}}\frac{H_i}{H_0}.
\eeq
Noting that $H_i \simeq ({\cal N} + 1) H_{\rm end}$ and substituting from
(\ref{eq:efold2}), (\ref{eq:tension}) \& (\ref{eq:temp0}) leads to
${\cal N} \simeq 70$.

\subsubsection{Post-inflationary evolution: The radiative regime}

We have demonstrated that the temperature of the universe
at the commencement of radiation domination is
a few GeV in exponential potential 
models with quantum-mechanical particle production (see also \cite{copeland}).
After a period during which $\rho_\phi$ can drop significantly below
$\rho_{\rm rad}$ (overshoot), 
the scalar field enters a `tracking regime' during
which the ratio $\rho_\phi/\rho_{\rm rad}$ is held constant
\cite{exp1,exp2}:
\beq
\frac{\rho_\phi}{\rho_{B} + \rho_\phi} = \frac{3(1 + w_B)}{\tilde\alpha^2}
\lleq 0.2,
\label{eq:tracker}
\eeq
equivalently
\beq
\frac{V''}{H^2} = \frac{9}{2}(1 - w_B)^2,
\eeq
($w_B = 0, ~1/3$ respectively for dust, radiation).
The inequality in (\ref{eq:tracker}) reflects nucleosynthesis constraints
which require the value of the dimensionless
parameter $\tilde\alpha$ to be large
$\tilde\alpha \ggeq 5$.

During tracking $w_\phi \simeq w_{\rm B}$, which leads to
\beq
\frac{1}{2} {\dot\phi}^2 = \frac{1+w_B}{1-w_B} V(\phi),
\label{eq:tracker0}
\eeq
substituting in (\ref{eq:tracker}) we get
\beq
\frac{{\dot\phi}^2}{\rho_{\rm total}} = \frac{3(1 + w_B)^2}{\tilde\alpha^2}.
\label{eq:tracker1}
\eeq
Since
\beq
\rho_{\rm total} = \frac{4M_P^2}{3(1+w_B)^2 t^2}
\label{eq:tracker2}
\eeq
we get
\beq
{\dot\phi}^2 = \frac{4M_P^2}{\tilde\alpha^2 t^2}
\label{eq:kinetic}
\label{eq:tracker3}
\eeq
{\em i.e.} the evolution of ${\dot\phi}$ is independent of the value of $w_B$ !
Integrating Eq.~(\ref{eq:kinetic}) between the epoch of radiation and
matter domination we get
\beq
\phi_{\rm MD} = \phi_{\rm RD} - \frac{2M_P}{\tilde\alpha}
\log{\left (\frac{t_{\rm MD}}{t_{\rm RD}}\right )}
\label{eq:tracker4}
\eeq
where $\phi_{\rm RD},\phi_{\rm MD}$ are scalar field values at the
commencement of radiative and matter dominated regime respectively.
It should be noted that
the inequality in (\ref{eq:tracker}) prevents the exponential potential
from becoming dominant during the course of cosmological evolution
and hence greatly limits its contribution to the `dark energy' during
later epochs. As we shall show in the next section, 
a small change in the potential can give rise to a model
of `quintessential inflation' \cite{pv99}
in which the scalar field density can drive the accelerated expansion of the
universe at the present time. 

\subsection{The cosine hyperbolic potential}
\label{sec:cosh}

The many excellent properties 
of the exponential potential also feature in the potential \cite{sw99}
\beq
V(\phi) = V_0(\cosh{\tilde\alpha\phi/M_P} - 1)^p, ~~ p > 0,
\label{eq:pot1}
\eeq
which has the following asymptotic forms
\ber
V(\phi) &\simeq& \frac{V_0}{2}e^{\tilde\alpha p\phi/M_P}, ~~ 
\tilde\alpha p\phi/M_P
\gg 1, ~\phi > 0,\label{eq:pot2}\\
V(\phi) &\simeq& \frac{V_0}{2}(\frac{\tilde\alpha\phi}{M_P})^{2p},
~~ |\tilde\alpha p\phi/M_P| \ll 1.
\label{eq:pot3}
\eer
An important difference between (\ref{eq:exponent}) and (\ref{eq:pot1}) is
that (\ref{eq:pot1}) goes over to the standard chaotic form for 
$p = 1$ and $\phi \lleq
M_P/\tilde\alpha$ allowing reheating to take place 
conventionally during oscillations of $\phi$ (provided $\phi$ couples
to other fields in nature). Reheating for
the exponential potential, which does not permit oscillations of $\phi$,
involves more exotic mechanisms \cite{exp2,copeland}.

By expanding eqn.~ (\ref{eq:pot1}) as
$V(\phi) = V_0\cosh{\tilde\alpha\phi/M_P} - V_0$ (for $p = 1$)
one finds that the
term $V_0$, which corresponds to a cosmological constant, has been subtracted
out in our potential. Since current observations favour the presence of a small
`$\Lambda$-term' \cite{sn,ss00} one might consider it appropriate
to include $V_0$ by replacing (\ref{eq:pot1})
by $V(\phi) = V_0\cosh{\tilde\alpha\phi/M_P}$.
From the discussion in the preceeding section it is easy to see that
such a model could give rise to inflation during an early epoch. It would
also give rise to dark energy today provided the value of $V_0$ were
set to \cite{sn,ss00} $V_0 \simeq 10^{-47} {\rm GeV}^4$.

\begin{figure}[h]
\centering
\resizebox{!}{2.5in}{\includegraphics{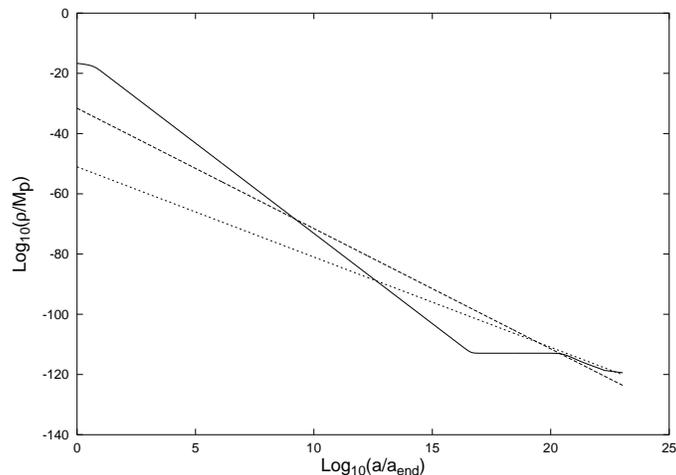}}
\caption{The post-inflationary energy density in 
the scalar field (solid line) radiation (dashed line) and cold dark
matter (dotted line) 
is shown as a function of the scale factor for the model 
decribed by (\ref{eq:pot1}) with 
$V_0 \simeq 5\times 10^{-46}$ GeV$^4$, $\tilde\alpha = 5$ and $p = 0.2$. 
The enormously large value of the scalar field
kinetic energy (relative to the potential)
ensures that the scalar field density overshoots the background
radiation value, after which $\rho_\phi$ remains approximately constant
for a substantially long period of time. At late times the scalar field briefly
tracks the background matter density before becoming dominant
and driving the current accelerated expansion of the universe. 
}
\label{fig:cosh1}
\end{figure}

\begin{figure}[h]
\centering
\resizebox{!}{2.5in}{\includegraphics{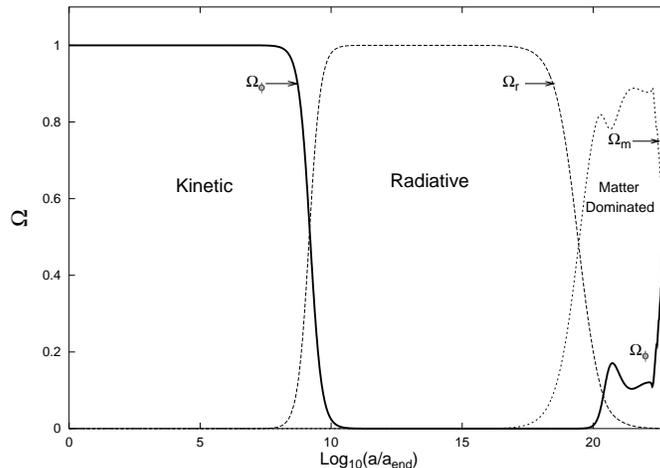}}
\caption{The dimensionless density parameter $\Omega$ is plotted as
a function of the scale factor for the model in figure \ref{fig:cosh1}.
Late time oscillations of the scalar field ensure that
the mean equation of state turns negative 
$\langle w_\phi\rangle \simeq -2/3$, giving rise to the current
epoch of cosmic acceleration with $a(t) \propto t^2$ and present day values
$\Omega_{0\phi} \simeq 0.7, \Omega_{0m} \simeq 0.3$.}
\label{fig:cosh2}
\end{figure}

Another means of getting a negative equation of state at late times
is by noting that, for small values of $\phi$, the form of the
potential changes to 
(\ref{eq:pot3}) giving rise to oscillations of the scalar field.
As demonstrated in \cite{sw99} scalar field
oscillations about $\phi = 0$ give rise to a mean
equation of state given by 
\beq
\langle w_\phi\rangle = \bigg\langle \frac{\frac{1}{2}{\dot\phi^2} - V(\phi)}
{\frac{1}{2}{\dot\phi^2} + V(\phi)} \bigg\rangle = \frac{p - 1}{p + 1}.
\label{eq:w}
\eeq
The resulting scalar field density and expansion factor have the form
\ber
&\langle\rho_\phi\rangle \propto a^{-3(1+\langle w_\phi\rangle)}
\label{eq:rho},\\
&a \propto t^c, ~ c = \frac{2}{3}(1+\langle w_\phi\rangle)^{-1}.
\label{eq:expansion}
\eer
We find that for $p=1$ the scalar field behaves like pressureless
(cold) dark matter. On the other hand $p < 1/2$ results in a negative mean
equation of state $\langle w_\phi\rangle < -1/3$, enabling the
scalar field to play the role of dark energy (quintessence).
We have numerically solved for the behaviour of this model after
including a radiative term (arising from inflationary paricle
production discussed in the previous section) and standard cold dark matter.
Our results for a particular realisation of the model with parameters
$V_0 \simeq 5\times 10^{-46}$ GeV$^4$, $\tilde\alpha = 5$ and $p = 0.2$
are shown in figures \ref{fig:cosh1} \& \ref{fig:cosh2}.
We find that, due to the very large value of the scalar kinetic energy
at the commencement of the radiative regime
(described by the ratio $\dot\phi^2/V(\phi)$), the scalar field density
overshoots the radiation density (see also \cite{stein}). 
After this, the value 
of $\rho_\phi$ stabilizes and 
remains relatively unchanged 
for a considerable length of time during which the scalar field
equation of state is $w_\phi \simeq -1$. 
Tracking commences late into
the matter dominated epoch and the universe accelerates today during rapid
oscillations of the scalar field. 
This model provides an interesting example of `quintessential
inflation'. However as shall discuss in section \ref{sec:gravity},
the long duration of the kinetic regime in this model results in a
large gravity wave background which could be in conflict with 
nucleosynthesis constraints.

\subsection{Power law potentials}

The calculations presented in section \ref{sec:exp}
can easily be extended to include other steep
potentials such as
\beq
V(\phi) =
\frac{(\mu M_4)^4}{\left ({\phi/M_4}\right )^{\tilde\alpha}}, ~~\tilde\alpha \gg 1
.
\label{eq:power0}
\eeq
The slow-roll parameter
in this case has the simple form
\beq
\epsilon \simeq \frac{\tilde\alpha^2}{4\pi}\left (\frac{\lambda_b}{\mu^4M_4^4}\right )
\left ( \frac{\phi}{M_4}\right )^{\tilde\alpha - 2}
\eeq
and one immediately notices that a larger value of the parameter
$\tilde\alpha$ assists slow roll
when $\phi/M_4 \ll 1$ provided
$\tilde\alpha > 2$. This behavior, which is completely at
odds with general relativistic
results, is clearly due to the presence of the brane correction
term $\lambda_b/V$ in
(\ref{eq:slow-roll}).

Inflation ends when $\epsilon \sim 1$ which leads to the useful result
\beq
\frac{V_{\rm end}}{\lambda_b} \left (\frac{\phi_{\rm end}}{M_4}\right )^2
\simeq \frac{\tilde\alpha^2}{4\pi}.
\label{eq:power1}
\eeq

The equation of motion in the slow roll regime
can be solved exactly resulting in
\beq
\phi^2(t) = \phi_i^2 + \frac{M_4\tilde\alpha}{\sqrt{3\pi\lambda_b}}(t-t_i).
\eeq

Next consider the expression for the number of inflationary e-foldings
\beq
{\cal N} \simeq - \frac{8\pi}{M_4^2}\int_{\phi_i}^{\phi_{\rm end}}\frac{V}{V'}
\left (\frac{V}{2\lambda_b}\right ) d\phi,
\label{eq:power2}
\eeq
for the power law potential (\ref{eq:power0}) one gets
\beq
{\cal N} \simeq \frac{4\pi}{\lambda_b}\frac{1}{\tilde\alpha(\tilde\alpha-1)}\left [
V_i\left (\frac{\phi_i}{M_4}\right )^2 -
V_{\rm end}\left (\frac{\phi_{\rm end}}{M_4}\right )^2 \right ].
\label{eq:power3}
\eeq
Equations (\ref{eq:power1}) \& (\ref{eq:power3}) lead to the following
useful relation
\beq
\frac{V_i}{\lambda_b}\left (\frac{\phi_i}{M_4}\right )^2 \simeq
\frac{1}{4\pi}\lbrack {\cal N}\tilde\alpha (\tilde\alpha - 2) + \tilde\alpha^2 \rbrack,
\label{eq:power4}
\eeq
where $\phi_i$ \& $V_i$ refer to the value of the scalar field 
and its potential at the commencement of inflation.
The amplitude of scalar perturbations in this model is \cite{maartens,copeland}
\ber
A_S^2 &=& \frac{64 \pi}{75 M_4^2} \left ( \frac{V_i}{V'}\right )^2
\left (\frac{V_i}{\lambda_b}\right )^3\left (\frac{V_i}{M_4^4}\right )\\
&=& \frac{64 \pi}{75 M_4^2} \left ( \frac{\phi_i}{\tilde\alpha}\right )^2
\left (\frac{V_i}{\lambda_b}\right )^3\left (\frac{V_i}{M_4^4}\right ),
\label{eq:power5}
\eer
which, after the substitution
\beq
\left ( \frac{\phi_i}{M_4}\right )^2 = \frac{1}{4\pi}
\frac{\lambda_b}{V_i} \lbrack {\cal N}\tilde\alpha (\tilde\alpha-2) + \tilde\alpha^2\rbrack
\label{eq:power6}
\eeq
acquires the form
\beq
A_s^2 \simeq \frac{16}{75}\left (\frac{V_i}{\lambda_b}\right )^2\left (\frac{V
_i}{M_4^4}\right ) \lbrack {\cal N} \frac{\tilde\alpha (\tilde\alpha - 2)}{\tilde\alpha^2} + 1
\rbrack.
\label{eq:power7}
\eeq
Using the COBE normalized value $A_s^2 \simeq 4\times 10^{-10}$ we get,
for $\tilde\alpha \gg 1$
\ber
V_i &\simeq& ({\cal N}+1) V_{\rm end},\\
\lambda_b &\simeq& \mu^\frac{24}{\tilde\alpha+4} M_4^4\left (
\frac{2\times 10^{-9}}{({\cal N}+1)^4}\right )^\frac{\tilde\alpha-2}{\tilde\alpha+4}
\left (\frac{4\pi}{\tilde\alpha^2}\right )^\frac{3\tilde\alpha}{\tilde\alpha+4},\\
V_{\rm end} &\simeq& \mu^\frac{16}{\tilde\alpha+4} M_4^4\left (
\frac{2\times 10^{-9}}{({\cal N}+1)^4}\right )^\frac{\tilde\alpha}{\tilde\alpha+4}
\left (\frac{4\pi}{\tilde\alpha^2}\right )^\frac{2\tilde\alpha}{\tilde\alpha+4},\\
\phi_{\rm end} &\simeq& \mu^\frac{4}{\tilde\alpha+4} M_4
\left (\frac{({\cal N}+1)^4}{2\times 10^{-9}}\right )
^\frac{1}{\tilde\alpha+4} \left (\frac{\tilde\alpha^2}{4\pi}\right)
^\frac{2}{\tilde\alpha+4}.
\label{eq:power8}
\eer
(We note that in the limit ${\tilde\alpha} \gg 1$, $V_{\rm end}/2\lambda_b \simeq {\tilde\alpha}^2/4\pi$, in agreement with our earlier results for the
exponential potential.)
The radiation density at this epoch can be obtained by substituting
for $V_{\rm end}, \lambda_b$ in
\beq
\left (\rho_r\right )_{\rm end} \simeq g_p T_{\rm end}^4
\simeq \frac{1}{5}g_p\frac{V_{\rm end}^4}{\lambda_b^2
M_4^4}
\label{eq:power_temp}
\eeq
which gives
\beq
T_{\rm end} \simeq \left (\frac{1}{5}\right )^{1/5}\mu^\frac{4}{\tilde\alpha+4}
\left (\frac{2\times 10^{-9}}{({\cal N} + 1)^4}\right )^\frac{\tilde\alpha+2}{2(\tilde\alpha+4)}
\left (\frac{4\pi}{\tilde\alpha^2}\right )^\frac{\tilde\alpha}{2(\tilde\alpha+4)}M_4.
\eeq
We also get
\beq
\left (\frac{\rho_\phi}{\rho_r}\right )_{\rm end}
\simeq \frac{5\lambda_b^2M_4^4}{g_p V_{\rm end}^3}
\simeq \frac{5}{2}\times 10^9 g_p^{-1}({\cal N}+1)^4
\label{eq:power_temp2}
\eeq
a result which is independent of both $\tilde\alpha$ and $\mu$ !

Consequently
\ber
&T_{\rm end}& \simeq \frac{2.6\times 10^{11}}{\tilde\alpha} \mu^{4/(\tilde\alpha+4)}~{\rm GeV}
~~ {\rm for}~ {\cal N} \sim 70,~\tilde\alpha \gg 1\\
&\left (\frac{\rho_\phi}{\rho_r}\right )_{\rm end}&
\simeq 6.4 \times 10^{16} g_p^{-1} ~~ {\rm for}~ {\cal N} \sim 70.
\label{eq:power_temp1}
\eer

Shortly after inflation has ended the kinetic regime commences
and the universe follows the expansion law (\ref{eq:evol_kin}).

Inverse power law potentials provide excellent models for quintessence
\cite{rp} and it is important to investigate whether, within the
braneworld framework, such potentials can describe {\em both}
quintessence and inflation (see also \cite{copeland,pow01}).
To achieve this one must ensure that the scalar field
remains in the kinetic regime for an appreciable length of time
($a/a_{\rm end} \sim 10^{10}$) long enough for the ratio $\rho_\phi/\rho_r$
to drop below unity and for radiation domination to commence.
During the kinetic regime the scalar field evolves as
\ber
\frac{\phi(t)}{M_4} &=& \frac{\phi_{\rm kin}}{M_4} +\sqrt{\frac{3}{4\pi}}
\log{\left({a(t) \over a_{\rm kin}}\right)}\nonumber\\
&=& \frac{\phi_{\rm kin}}{M_4} +\sqrt{\frac{1}{12\pi}}
\log{\left({t \over t_{\rm kin}}\right)}
\label{eq:kin01}
\eer
where $\phi_{\rm kin} > \phi_{\rm end}$ and from (\ref{eq:power_temp2})
we see that $\phi_{\rm end}/M_4 \ggeq 1$. The equalization of
the density in stiff-matter and radiation marks the commencement
of the radiative regime and the value of the scalar field at
this juncture is given by
\beq
\frac{\phi_{\rm eq}}{M_4} = \frac{\phi_{\rm kin}}{M_4}
+ \sqrt{\frac{3}{16\pi}}
\log{\left(\frac{\rho_\phi}{\rho_r}\right )_{\rm kin}}.
\label{eq:pow_motion}
\eeq
A lower limit on
$\left (\rho_\phi/\rho_r\right )_{\rm kin}$
can be derived by assuming that the ratio $\rho_\phi/\rho_r$
decreases as $a^{-2}$ between the end of inflation and the
commencement of the kinetic regime (in fact it decreases slower),
in addition if we make the conservative assumption
$T_{\rm kin} \simeq 0.01 T_{\rm end}$ we get
\beq
\left (\frac{\rho_\phi}{\rho_r}\right )_{\rm kin} =
\left (\frac{T_{\rm end}}{T_{\rm kin}}\right )^{-2}
\left (\frac{\rho_\phi}{\rho_r}\right )_{\rm end} \ggeq 6\times 10^{12}
g_p^{-1}.
\eeq
Eqn.~
(\ref{eq:pow_motion}) can now be used to obtain a lower bound
on the scalar field value at the commencement of the radiative
regime
\beq
\frac{\phi_{\rm eq}}{M_4} \ggeq \frac{\phi_{\rm kin}}{M_4} +
\sqrt{\frac{3}{16\pi}}
\log{\left(6\times 10^{12}g_p^{-1} \right )}. 
\eeq
A large value of $\phi_{\rm eq}$
indicates that the field has rolled down to
regions where the potential is less steep. During tracking
flat potentials can give rise to inflation, therefore
one must ensure that inflation does not recur before the universe becomes
radiation dominated.
Imposing the requirement $\epsilon_{\rm FRW} > 1$ where
\beq
\epsilon_{\rm FRW} = \frac{M_4^2}{16\pi}\left (\frac{V'}{V}\right )^2
= \frac{{\tilde\alpha}^2}
{16\pi}\left (\frac{\phi}{M_4}\right )^{-2}
\eeq
is the slow roll parameter, we get
\beq
\frac{\phi}{M_4} \lleq \frac{\phi_*}{M_4} \equiv
\frac{\tilde\alpha}{\sqrt{16\pi}}.
\label{eq:slow_roll}
\eeq

\begin{figure}[h]
\centering
\resizebox{!}{2.5in}{\includegraphics{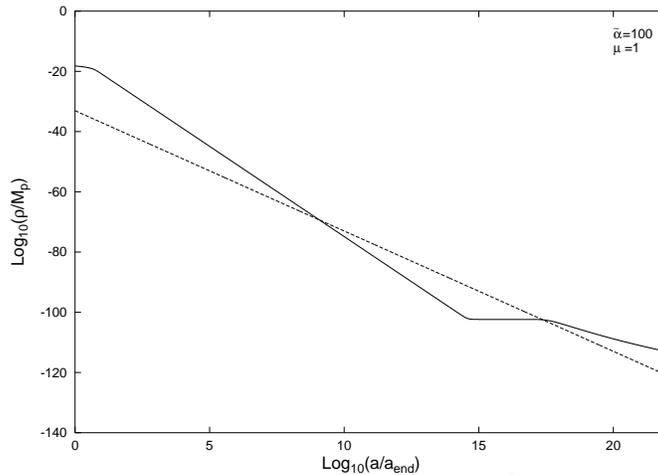}}
\caption{The post-inflationary evolution of the scalar field density 
(solid line)
is shown for the 
potential (\ref{eq:power0}) with $\mu = 1$ and
$\tilde\alpha = 100$. The radiation density is also shown (dashed line).
We see that the scalar field energy density dominates the expansion dynamics
of the universe during both early and late times. This model re-inflates
much too soon, resulting in an unacceptably large value of the dark 
energy today.
}
\label{Den100}
\end{figure}


\begin{figure}[h]
\centering
\resizebox{!}{2.5in}{\includegraphics{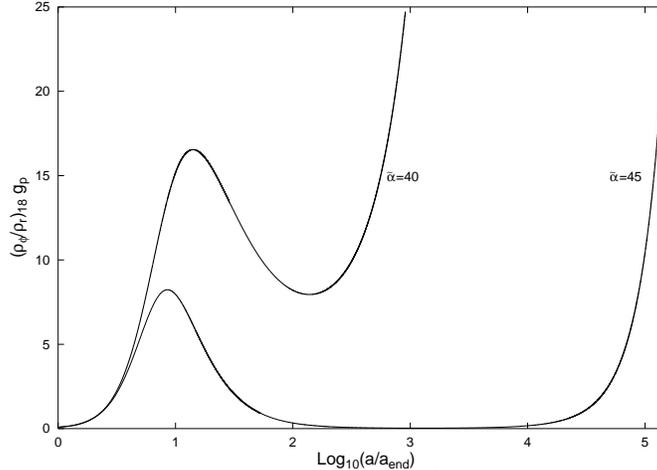}}
\caption{Post-inflationary evolution of $(\rho_{\phi}/\rho_{r})_{18}=
(\rho_{\phi}/\rho_{r})\times10^{-18}$ is shown
for the inverse power law potential (\ref{eq:power0}) 
$V(\phi) \propto \phi^{-{\tilde\alpha}} ~(\mu = 1)$.
The initial increase in $\rho_{\phi}/\rho_{r}$ is due to brane corrections
immediately after inflation. During the kinetic regime
$\rho_{\phi}/\rho_{r}$ initially decreases as $a^{-2}$ but increases again
during a second epoch of inflation which
commences at $a/a_{end} \sim 10^{2}, 10^{4}$ for ${\tilde\alpha} = 40, 45$
respectively. In both cases the post-inflationary 
expansion of
the universe is {\em always} scalar field dominated and the universe 
{\em never} enters the radiative regime.
}
\label{Den40}
\end{figure}

\begin{figure}[h]
\centering
\resizebox{!}{2.5in}{\includegraphics{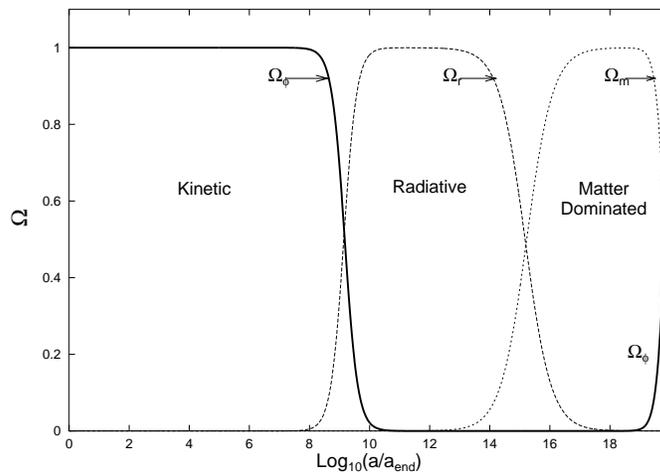}}
\caption{Post-inflationary evolution of the density parameter 
$\Omega$ is shown
for: (i) the inverse power law potential (\ref{eq:power0}) with 
$\tilde{\alpha} = 20, \mu \simeq 10^{-26}$ (solid line) 
(ii) radiation
(dashed line) and (iii) matter (dotted line). 
In this model of `quintessential inflation'
the scalar field dominates the energy density
of the universe twice, initially during inflation and finally during the
current `dark energy' dominated epoch, giving present day values
$\Omega_{0\phi} \simeq 0.7, \Omega_{0m} \simeq 0.3$.
}
\label{Lidden}
\end{figure}

The estimate given by (\ref{eq:slow_roll}) is valid 
provided the scalar field enters
the tracking regime soon after radiation domination. Numerical estimates
show that this is not always the case. Indeed, as demonstrated in 
Figure \ref{Den100}, 
the scalar field can remain in the kinetic regime for a 
considerable period of time after the universe becomes radiation dominated.
For large values of $\mu \sim O(1)$ and ${\tilde\alpha} \lleq 75$
we find that the scalar field can cause the universe to re-inflate
either before or soon
after the commencement of the radiative regime (see figures \ref{Den40}
\& \ref{Den100}). 
As a result, the matter dominated regime is {\em never} reached. 
However, as demonstrated in \cite{pow01}, for a narrow range of 
parameter values the universe can be made to inflate at the present epoch 
providing a possible model for `quintessential inflation'. 
In figure \ref{Lidden} 
we show the evolution of the universe filled with matter, 
radiation and a scalar field 
potential with parameters
${\tilde\alpha} = 20$, $\mu \simeq 10^{-26}$. The scalar field in this case
successfully plays the dual role of being the inflaton at an early epoch
and dark energy today, lending support to the analysis of \cite{pow01}
(see also \cite{pv99,powerlaw,maeda}).
From (\ref{eq:power_temp2}) it is easy to show that the 
duration of the kinetic regime, described by the
ratio $a_{\rm eq}/a_{\rm kin} = 
T_{\rm kin}/T_{\rm eq} \sim 10^9 g_p^{-1/2}$,
depends very weakly upon the values of $\mu$ and $\tilde{\alpha}$.
For the cosmologically relevant values $\tilde{\alpha} = 20, 
\mu  \simeq 10^{-26}$ we find
$T_{\rm kin} \sim 10^6 {\rm GeV},
T_{\rm eq} \sim 10 {\rm MeV}$ and $a_{\rm eq}/a_{\rm kin} \sim
10^8$.
As we show in the next section, relic gravity waves
created during inflation impose strong constraints
on braneworld inflationary
models having kinetic regimes of such long duration.

\begin{figure}[h]
\centering
\resizebox{!}{2.5in}{\includegraphics{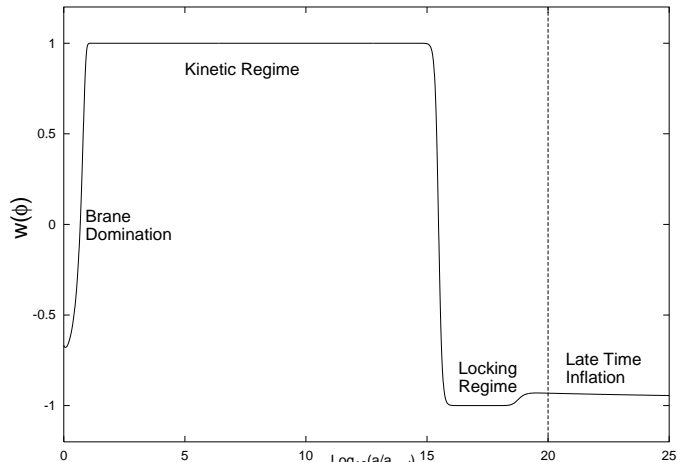}}
\caption{Post-inflationary evolution of the equation of state 
of the scalar field model
considered in figure \ref{Lidden}.
The evolution of $w(\phi)$
follows four distinct
stages: (i) {\em Post-inflationary brane domination.}
Diminishing post-inflationary brane effects cause
the equation of state in the $\phi$-field to gradually increase from
$w(\phi) \simeq -2/3$ to $w(\phi) \simeq 1$.
(ii) {\em The kinetic regime}. During this regime
the kinetic energy of the scalar field exceeds
its potential energy leading to $w(\phi) \simeq 1$.
(iii) {\em Locking}. 
During the radiative regime the energy density in the scalar
field overshoots the energy density in radiation resulting 
in an extensive period during which the 
scalar field
equation of state locks to $w(\phi) \simeq -1$. The scalar field during this
regime behaves like an effective cosmological constant since its
energy density $\rho_\phi$ does not change appreciably with time.
(iv) {\em Late time inflation}. During the late matter dominated regime
the ratio of the
scalar field density to that in matter
grows until the scalar field dominates the energy density of the universe. 
This leads to the current epoch of accelerated expansion 
(`late time inflation')
during which $w(\phi) \lleq -0.9$. 
The vertical dashed line corresponds to the present epoch.
}
\label{Lidden1}
\end{figure}

\section{Gravity wave background}
\label{sec:gravity}
An important feature of inflationary scenarios of the very early universe
is the quantum
mechanical production of relic gravity waves \cite{star,allen,sahni}
which create fluctuations in the cosmic microwave background and whose
stochastic signature presents a challenge to gravity wave observatories such
as LIGO and LISA.
Gravity waves in a spatially homogeneous and isotropic background geometry
satisfy the minimally coupled Klein-Gordon equation $\sq h_{ik} = 0$ \cite{grish},
which, after a separation of variables 
$h_{ij} = \phi_k(\tau) e^{-i{\bf k}{\bf x}}e_{ij}$
($e_{ij}$ is  the polarization tensor) reduces to
\beq
{\ddot \phi_k} + 2\frac{\dot a}{a}{\dot \phi}_k + k^2\phi_k = 0
\label{eq:kg2}
\eeq
where $\tau = \int dt/a(t)$ is the conformal time coordinate
and $k = 2\pi a/\lambda$ is the comoving wavenumber.
Since brane driven inflation is near-exponential we can write
$a = \tau_0/\tau$ $(\vert\tau\vert < \vert\tau_0\vert)$, in this case
normalized positive frequency solutions of (\ref{eq:kg2}) 
corresponding to the adiabatic
vacuum in the `in state' are given by \cite{bd78,lmw00}
\beq
\phi_{\rm in}^+(k,\tau) = \left (\frac{\pi \tau_0}{4}\right )^{1/2}
\left (\frac{\tau}{\tau_0}\right )^{3/2}
H_{3/2}^{(2)}(k\tau)F(H_{\rm in}/\tilde\mu)
\label{eq:hankel}
\eeq
where $\tilde\mu = M_5^3/M_4^2$ and $H_{\rm in} \equiv -1/\tau_0$ is the
inflationary Hubble parameter.
The term
\beq
F(x) = \left (\sqrt{1+x^2} - x^2\log{\lbrace \frac{1}{x} + \sqrt{1 + \frac{1}{x^2}}\rbrace }
\right )^{-1/2}
\eeq
is responsible for the increased gravity wave amplitude
in braneworld inflation \cite{lmw00}.
At low energies ($H_{\rm in}/{\tilde\mu} \ll 1 \Rightarrow
\rho/\lambda_b \ll 1$) $F \simeq 1$,
while at high energies ($\rho/\lambda_b \gg 1$)
\beq
F \simeq \sqrt{\frac{3H_{\rm in}}{2\tilde\mu}} \simeq
\sqrt{\frac{3\rho_{\rm in}}{2\lambda_b}}.
\eeq
The corresponding dimensionless amplitude of gravity waves is given by
\beq
h_{\rm GW}^2 = \left (\frac{H_{\rm in}}{M_4}\right )^2 F^2
\simeq
\left (\frac{H_{\rm in}}{M_4}\right )^2 \left (\frac{3 \rho}{2\lambda_b}\right )
.
\label{eq:gw}
\eeq

For the scalar field models discussed earlier, Eqn. ~(\ref{eq:gw}) reduces to
\beq
h_{\rm GW}^2 \simeq 2\pi\frac{V_{\rm HC}^3}{\lambda_b^2 M_4^4},
\label{eq:gw01}
\eeq
where $V_{\rm HC}$ corresponds to the value of the potential when the 
given gravity wave mode left the Hubble radius during inflation.
From (\ref{eq:power7}) \& (\ref{eq:gw01})
we find a simple relation between $h_{\rm GW}$ and $A_s$
\beq
h_{\rm GW}^2  = \frac{75\pi}{8}A_s^2\left [{\cal N}\frac{(\tilde\alpha-2)}{\tilde\alpha}
+ 1\right ]^{-1}
\label{eq:gw02}
\eeq
which reduces to
\beq
h_{\rm GW}^2 \simeq \frac{75 \pi}{8}\frac{A_s^2}{{\cal N} + 1}
\label{eq:gw03}
\eeq
for $\tilde\alpha \gg 1$. It is easy to show that (\ref{eq:gw03})
is also the correct expression
for the gravity wave amplitude in inflationary models with exponential
potentials
for {\em arbitrary values} of the parameter $\tilde\alpha$
defined in (\ref{eq:exponent}).
Finally, using the COBE-normalized values $A_s \simeq 2\times 10^{-5}$ and
assuming ${\cal N}_{\rm total} \simeq 70$ we obtain
\beq
h_{\rm GW}^2 \simeq 1.7 \times 10^{-10}
\label{eq:hubble}
\eeq
for the exponential and inverse power law potentials discussed in the
preceding sections. (In deriving (\ref{eq:hubble}) we have used the
result that most of the COBE signal is produced by scalar density
perturbations \cite{lmw00,copeland}.)

The `out state' is described by a linear superposition of positive and negative
frequency solutions to (\ref{eq:kg2}). For power law expansion
$a = (t/t_0)^p \equiv (\tau/\tau_0)^{1/2-\mu}$, we have
\beq
\phi_{\rm out}(k,\tau) = \alpha \phi^{(+)}_{\rm out}(k\tau) + 
\beta \phi^{(-)}_{\rm out}(k\tau)
\label{eq:hankel1}
\eeq
where
\beq
\phi^{(+,-)}_{\rm out}(k\tau) = \left (\frac{\pi \tau_0}{4}\right )^{1/2}
\left (\frac{\tau}{\tau_0}\right )^\mu H_{\vert\mu\vert}^{(2,1)}(k\tau),
\eeq
and $\mu$ is related to the post-inflationary
equation of state
\beq
\mu = \frac{3}{2}\left (\frac{w - 1}{3 w + 1}\right ).
\eeq
The Wronskian normalization condition $W_\tau(\phi_{\rm out},\phi_{\rm out}^*)
= i/a^2$ ensures that $|\alpha^2| - |\beta^2| = 1$.
The energy density of relic gravity waves is given by \cite{fp77a,fp77b}
\beq
\epsilon_{\rm g} = \langle T_0^0\rangle = \frac{1}{2\pi^2 a^2}
\int{dk k^2 \left ( \vert {\dot\phi}_{\rm out}\vert ^2
+ k^2\vert \phi_{\rm out}\vert^2\right )},
\eeq
which reduces to
\beq
\epsilon_{\rm g} = \frac{1}{\pi^2 a^4}\int
dk k^3 \vert\beta\vert^2,
\label{eq:gw_energy}
\eeq
if modes within the horizon provide the dominant contribution to
the energy density. 
The corresponding spectral energy density is simply
\beq
\tilde\epsilon_{\rm g}(k) \equiv \frac{d}{d\log{k}}\epsilon_{\rm g}
= \frac{1}{\pi^2a^4}k^4\vert\beta(k)\vert^2.
\label{eq:spectrum}
\eeq

The Bogoliubov coefficients $\alpha$ \& $\beta$ can be determined after imposing
suitable junction conditions on $\phi_{\rm in}$ and $\phi_{\rm out}$
as discussed in detail in \cite{grish,star,allen,sahni}.
For instance the small argument limit of the Hankel function
\beq
H_\mu ^{(2,1)} (k\tau)  {\mathrel{\mathop{=}_{k\tau \ll 2\pi} } }
\phantom{...}{({k\tau \over 2})^\mu \over \Gamma
(1+\mu)} \pm {i\over \pi} \Gamma (\mu) \left({k \tau \over 2}\right) ^{-\mu}
\,\, (\mu \neq 0)
\eeq
\beq
H_0^{(2,1)} (k\tau)  {\mathrel{\mathop{=}_{k\tau \ll 2\pi} } }
\phantom{...} 1 \mp \frac{2i}{\pi} \log{(k\tau)}
\eeq
combined with the observation that the amplitude of a
general solution to the wave equation (\ref{eq:kg2})
freezes to a constant value on scales larger
than the cosmological horizon
\beq
\phi (k\tau) {\mathrel{\mathop{=}_{k\tau \ll 2\pi} } }
A(k) + B(k) \int {d(\tau/\tau_0) \over a^2}
\eeq
allows us to match the `in' and `out' modes (\ref{eq:hankel}) \&
(\ref{eq:hankel1})
and determine
Bogoliubov coefficients $\alpha$ \& $\beta$.

Following this prescription (which has been described in detail in \cite{sahni})
 we proceed to evaluate the spectral density of gravity waves created
in inflationary braneworld models.
Before commencing on a detailed analysis of the problem let us first consider
the simple but illustrative case of a universe which, after inflating,
enters a non-inflationary regime with equation of state
$0 \leq w < 1 ~(-3/2 \leq \mu < 0$). The Bogoliubov coefficients in this case
have been found to be \cite{sahni}
\ber
\alpha &=& \frac{i}{2}\left\lbrack \gamma
\left (\frac{k\tau_0}{2}\right )^{-(3/2 + |\mu|)} + \gamma^{-1}
\left (\frac{k\tau_0}{2}\right )^{3/2 + |\mu|} \right\rbrack\nonumber\\
\beta &=& \frac{i}{2}\left\lbrack \gamma
\left (\frac{k\tau_0}{2}\right )^{-(3/2 + |\mu|)} - \gamma^{-1}
\left (\frac{k\tau_0}{2}\right )^{3/2 + |\mu|} \right\rbrack,
\label{eq:bog0}
\eer
for $k\tau_0 < 2\pi$, here $\gamma = \Gamma (1+|\mu|)/2\sqrt{\pi}$.
(On smaller than
horizon scales
($k\tau > 2\pi$) the adiabatic theorem gives
$\alpha \simeq 1, \beta \simeq 0$ \cite{bd82,allen}.)

From (\ref{eq:gw_energy}) \& (\ref{eq:bog0}) it is easy to show that
\ber
\epsilon_{\rm g} &\propto& a^{-4} \,\, {\rm for} \,\, w > 1/3\nonumber\\
\epsilon_{\rm g} &\propto& \epsilon_{\rm B} \,\, {\rm for} \,\, w < 1/3,
\eer
where $\epsilon_{\rm B}$ is the background matter density.
In other words the gravity wave energy density scales as radiation
if the equation of state of background matter driving the expansion of the
universe is
$P_{\rm B} > \epsilon_{\rm B}/3$. In the reverse case when
$P_{\rm B} < \epsilon_{\rm B}/3$, the gravity wave equation of state
mimicks that of the background, so that
$\epsilon_{\rm g}/\epsilon_{\rm B} \simeq {\rm constant}$.
This `tracker-like' behavior of the
gravity wave energy density was first discovered in
\cite{allen,sahni}. (For $P_{\rm B} = \epsilon_{\rm B}/3$
$\epsilon_{\rm g} \propto a^{-4} \log{(\tau/\tau_0)}$.)

From (\ref{eq:spectrum}) we also find that \cite{sahni}
\beq
\tilde\epsilon_{\rm g} \propto k^{1-2|\mu|},
\label{eq:spectra}
\eeq
as a result, gravity waves will have
have a blue spectrum for equations of state stiffer than radiation
($w_\phi > 1/3, |\mu| < 1/2$)
and a red spectrum if $w_\phi < 1/3, ~|\mu| > 1/2$.
Our current cosmological model passes through three post-inflationary
expansion epochs
during which its equation of state is succesively:
stiff ($w \simeq 0$), radiative ($w \simeq 1/3$), matter-dominated
($w \simeq 0$). Gravity waves created during these separate epochs are
therefore likely to have a scale-dependent tilt which will vary from
being `blue' during the kinetic regime to
`white' ($\equiv$ flat) during radiation domination to `red' during
matter-domination \cite{sahni,giovan}. 
We shall now demonstrate this explicitly.
According to a general result which allows us to determine
Bogoliubov coefficients
in a multi-component universe,
the Bogoliubov coefficients ${\cal B}_{n}$
describing particle production during the `n-th' successive epoch
can be determined from
\cite{allen,sahni}
\beq
{\cal B}_n = {\cal B}_{1\rightarrow 2}
{\cal B}_{2\rightarrow 3}\cdot\cdot\cdot {\cal B}_{n-1 \rightarrow n}
\eeq
where
\beq
{\cal B} = \left(\matrix{\alpha&\beta\cr
\beta^*&\alpha^*}\right).
\eeq
In the present cosmological model the universe passes through four
expansion stages: Inflation (1) $\to$ kinetic regime (2) $\to $
radiative regime (3) $\to$ matter-dominated regime (4).
Bogoliubov coefficients corresponding to each of these regimes are obtained
below assuming that the transition from one regime to the next is
instantaneous.

1. {\em Kinetic regime}. Bogoliubov coefficients corresponding to modes
created in the course
of the `Inflation $\to$ kinetic' ($1 \to 2$) transition are given by \cite{sahni}
\ber
\alpha_{1 \to 2} \equiv \alpha_{\rm kin} &=& \frac{i}{2}\left\lbrack \frac{1}{2\sqrt{\pi}}
\left (\frac{k\tau_{\rm kin}}{2}\right )^{-3/2} + 2\sqrt{\pi}
\left\lbrace 1 + \frac{2i}{\pi}\log (k\tau_{\rm kin})\right\rbrace
\left (\frac{k\tau_{\rm kin}}{2}\right )^{3/2}\right\rbrack\nonumber\\
\beta_{1 \to 2} \equiv \beta_{\rm kin} &=& \frac{i}{2}\left\lbrack \frac{1}{2\sqrt{\pi}}
\left (\frac{k\tau_{\rm kin}}{2}\right )^{-3/2} - 2\sqrt{\pi}
\left\lbrace 1 - \frac{2i}{\pi}\log (k\tau_{\rm kin})\right\rbrace
\left (\frac{k\tau_{\rm kin}}{2}\right )^{3/2}\right\rbrack,
\label{eq:bog1}
\eer
where $\tau_{\rm kin}$ corresponds to the commencement of the
kinetic regime.

2. {\em Radiative regime}. Bogoliubov coefficients describing
modes created during this regime arise due to successive transitions:
Inflation $\to$ kinetic ($1 \to 2$) and kinetic $\to$ radiative ($2 \to 3$).
As a result
\beq
{\cal B}_{\rm RD} = {\cal B}_{1 \to 2}\cdot {\cal B}_{2 \to 3}
\label{eq:bog2}
\eeq
where the Bogoliubov coefficients ${\cal B}_{1 \to 2}$
were derived in 
(\ref{eq:bog1}).
Particle production during the `kinetic $\to$ radiative'
transition is described by
\ber
\alpha_{2 \to 3} &=& \frac{1}{2}\left\lbrack\frac{\sqrt{\pi}}{2}
\left (\frac{k\tau_{\rm RD}}{2}\right )^{-1/2} +
\frac{2}{\sqrt{\pi}}\left (\frac{k\tau_{\rm RD}}{2}\right )^{1/2}
\right\rbrack\nonumber\\
\beta_{2 \to 3} &=& \frac{1}{2}\left\lbrack\frac{\sqrt{\pi}}{2}
\left (\frac{k\tau_{\rm RD}}{2}\right )^{-1/2} -
\frac{2}{\sqrt{\pi}}\left (\frac{k\tau_{\rm RD}}{2}\right )^{1/2}
\right\rbrack,
\label{eq:bog3}
\eer
where $\tau_{\rm RD}$ corresponds to the commencement of the radiative regime.
From (\ref{eq:bog2}), (\ref{eq:bog1}) \& (\ref{eq:bog3}) we finally obtain
\ber
\alpha_{\rm RD} &=& \frac{i}{2}\left\lbrack \frac{1}{k^2\tau_{\rm kin}^{3/2}
\tau_{\rm RD}^{1/2}} + k^2\tau_{\rm kin}^{3/2}\tau_{\rm RD}^{1/2} +
ik\frac{\tau_{\rm kin}^{3/2}}{\tau_{\rm RD}^{1/2}}\log {(k\tau_{\rm kin})}
\right\rbrack\nonumber\\
\beta_{\rm RD} &=& \frac{i}{2}\left\lbrack \frac{1}{k^2\tau_{\rm kin}^{3/2}
\tau_{\rm RD}^{1/2}} - k^2\tau_{\rm kin}^{3/2}\tau_{\rm RD}^{1/2} +
ik\frac{\tau_{\rm kin}^{3/2}}{\tau_{\rm RD}^{1/2}}\log {(k\tau_{\rm kin})}
\right\rbrack.
\label{eq:bog4}
\eer

3. {\em Matter dominated regime.}
Gravity waves created during the matter-dominated regime
bear the imprint of three successive
transitions:
Inflation (1) $\to$ kinetic (2), kinetic (2) $\to$ radiative (3),
radiative (3) $\to$ matter-dominated (4). The Bogoliubov coefficients
are therefore determined from
\beq
{\cal B}_{\rm MD} = {\cal B}_{1 \to 2}\cdot {\cal B}_{2 \to 3}
\cdot {\cal B}_{3 \to 4}
\label{eq:bog5}
\eeq
where
\ber
\alpha_{3 \to 4} &=& \frac{1}{2}\left\lbrack\frac{3}{k\tau_{\rm MD}}
+ \frac{k\tau_{\rm MD}}{3}\right\rbrack\nonumber\\
\beta_{3 \to 4} &=& \frac{1}{2}\left\lbrack\frac{3}{k\tau_{\rm MD}}
- \frac{k\tau_{\rm MD}}{3}\right\rbrack
\label{eq:bog6}
\eer
and $\tau_{\rm MD}$ refers to the commencement of the current matter
-dominated regime. From (\ref{eq:bog5}), (\ref{eq:bog1}), (\ref{eq:bog3})
\& (\ref{eq:bog6}) we finally obtain
\ber
\alpha_{\rm MD} &=& \frac{i}{2}\left\lbrack \frac{3}{k^3\tau_{\rm MD}
\tau_{\rm RD}^{1/2}\tau_{\rm kin}^{3/2}} + 3i\frac{\tau_{\rm kin}^{3/2}}
{\tau_{\rm MD}\tau_{\rm RD}^{1/2}}\log{(k\tau_{\rm kin})} +
\frac{k^3\tau_{\rm MD}\tau_{\rm RD}^{1/2}\tau_{\rm kin}^{3/2}}{3}
\right\rbrack\nonumber\\
\beta_{\rm MD} &=& \frac{i}{2}\left\lbrack \frac{3}{k^3\tau_{\rm MD}
\tau_{\rm RD}^{1/2}\tau_{\rm kin}^{3/2}} + 3i\frac{\tau_{\rm kin}^{3/2}}
{\tau_{\rm MD}\tau_{\rm RD}^{1/2}}\log{(k\tau_{\rm kin})} -
\frac{k^3\tau_{\rm MD}\tau_{\rm RD}^{1/2}\tau_{\rm kin}^{3/2}}{3}
\right\rbrack.
\label{eq:bog7}
\eer
It is easy to show that the Bogoliubov coefficients evaluated in
(\ref{eq:bog1}), (\ref{eq:bog4}) \& (\ref{eq:bog7})
satisfy $|\alpha_{\rm kin}|^2 - |\beta_{\rm kin}|^2 =
|\alpha_{\rm RD}|^2 - |\beta_{\rm RD}|^2 =
|\alpha_{\rm MD}|^2 - |\beta_{\rm MD}|^2 = 1$. Inspecting $\alpha$ \& $\beta$
in (\ref{eq:bog1}), (\ref{eq:bog4}) \& (\ref{eq:bog7})
one finds that the dominant role in each of these expressions
is played by the first term, so that effectively

\ber
\vert \beta_{\rm kin}\vert^2 &\simeq& \frac{1}{2\pi} (k\tau_{\rm kin})^{-3},
~~ 2\pi\tau_{\rm kin}^{-1} < k \leq k_{\rm RD},\label{eq:bog8}\\
\vert \beta_{\rm RD}\vert^2 &\simeq& \frac{1}{4}
k^{-4}\tau_{\rm RD}^{-1}\tau_{\rm kin}^{-3},
~~ k_{\rm MD} \leq k < k_{\rm RD},\label{eq:bog9}\\
\vert \beta_{\rm MD}\vert^2 &\simeq& \frac{9}{4}k^{-6}\tau_{\rm MD}^{-2}
\tau_{\rm RD}^{-1}\tau_{\rm kin}^{-3},
~~ 2\pi\tau^{-1} \lleq k < k_{\rm MD}
\label{eq:bog10}
\eer
where $k\tau = 2\pi$ describes the comoving horizon scale and
$k_{\rm RD} = \pi/2\tau_{\rm RD}$, $k_{\rm MD} =
3/\tau_{\rm MD}$ have been obtained using junction conditions.

We are now in a position to evaluate the spectral energy density of
relic gravity waves given in (\ref{eq:spectrum}). From (\ref{eq:spectrum}) and
Eqns.~(\ref{eq:bog8}) -- (\ref{eq:bog10}) we find that the spectral density
of relic gravity waves is very sensitive to the equation of state
of matter driving the expansion of the universe.
Our brane-inspired cosmological model has
three distinct \& lengthy
post-inflationary epochs during which the effective
equation of state
is successively:
(i) stiff ($w = 1$), (ii) radiation dominated ($w = 1/3$)
and (iii) matter dominated ($w = 0$).
Reflecting this, the relic
gravity wave spectrum
will have three main components:
$\tilde\epsilon_{\rm g} \propto \lambda^{-1}$ for $\lambda < \lambda_{\rm RD}$,
$\tilde\epsilon_{\rm g} = {\rm constant}$ for $\lambda_{\rm RD} \lleq
\lambda < \lambda_{\rm MD}$ and
$\tilde\epsilon_{\rm g} \propto \lambda^{-2}$
for $\lambda_{\rm MD}^{\rm h} \lleq \lambda < \lambda_0^{\rm h}$;
$\lambda_{\rm RD}^{\rm h}$ \& $\lambda_{\rm MD}^{\rm h}$ are related
to the comoving
hubble radius 
at the commencement of the radiation dominated and matter dominated regimes
respectively (precise values of these quantities will be given later)
while $\lambda_0^{\rm h} \simeq 10^{28}$ cm is the present distance
to the cosmological horizon.

From (\ref{eq:spectrum}) \& (\ref{eq:bog8})
we find that a long kinetic regime strongly affects the gravity
wave spectrum giving it considerable power on short wavelength scales
$\tilde\epsilon_g(k) \propto \lambda^{-1}$.
Since gravity waves corresponding to this
part of the spectrum
were created prior to the radiative regime
we must ensure that the
integrated gravity wave energy density satisfies the stringent constraints
set by cosmological nucleosynthesis. 
To resolve this issue we examine Eq. (\ref{eq:gw_energy})
which after the substitution $\beta \to \beta_{\rm kin}$ reduces to
\beq
\epsilon_{\rm g} = \frac{32}{3\pi} h_{\rm GW}^2 \epsilon_{\rm B}
\left (\tau/\tau_{\rm kin}\right )
\label{eq:gw_energy1}
\eeq
where $\epsilon_{\rm B}$ is the background density of stiff scalar matter
driving the expansion of the universe.
The radiative regime is reached when
$\tau = \tau_{\rm eq}$ and $\epsilon_{\rm B} = \epsilon_{\rm stiff} +
\epsilon_{\rm rad} \simeq 2\epsilon_{\rm rad}$,
since
$\left (\tau_{\rm eq}/\tau_{\rm kin}\right ) = \left (T_{\rm kin}/
T_{\rm eq}\right )^2$, we arrive at the relationship
\beq
\epsilon_{\rm g}(\tau = \tau_{\rm eq})
= \frac{64}{3\pi} h_{\rm GW}^2 \epsilon_{\rm rad}(\tau = \tau_{\rm eq})
\left (\frac{T_{\rm kin}}{T_{\rm eq}}\right )^2.
\eeq
Substituting from (\ref{eq:temp1}), (\ref{eq:temp2}) \& (\ref{eq:hubble})
and allowing for the fact that the Hubble crossing
gravity wave amplitude is somewhat
smaller at $\tau_{\rm eq}$ than it is at present, we get
\beq
\epsilon_{\rm g} \sim 2\times 10^{8} g_p^{-1}\epsilon_{\rm rad}
\label{eq:gw_constraints}
\eeq
{\em i.e.} unless $g_p > 10^{9}$
the energy density in gravity waves will exceed
the radiation density
by a very substantial amount violating nucleosynthesis
constraints which demand $\epsilon_{\rm g} \lleq 0.2 \epsilon_{\rm rad}$.
(Eqn (\ref{eq:gw_constraints}) was derived under the assumption that
$\tilde\alpha = 3$, larger values of $\tilde\alpha$ will increase
$\epsilon_{\rm g}$ and further exacerbate the situation.)
From (\ref{eq:gw_energy1}) it also follows that the gravity wave energy
density will begin to dominate scalar field matter when the
universe has expanded by a factor $a_{\rm eq}/a_{\rm kin} \sim 10^5$.
The backreaction of gravity waves will at this point effectively end 
the kinetic regime and the
expansion of the universe will change from
$a \propto \tau^{1/2}$ characteristic of the stiff-matter regime,
to $a \propto \tau$ characteristic of the radiative regime (see also
\cite{giovan}).
A similar result is obtained for the inverse power law potentials
discussed in the previous section.

It is interesting that the equations (\ref{eq:ratio_end}),
(\ref{eq:gw01}) \& (\ref{eq:gw_energy1}) can also be used to obtain a
{\em model independent} constraint on brane-induced quintessential
inflation in which reheating arises solely on account of inflationary particle production. Indeed, since $(\tau_{\rm eq}/\tau_{\rm kin}) =
(a_{\rm eq}/a_{\rm kin})^2 = (\rho_\phi/\rho_{\rm rad})_{\rm kin}$
we find, using the relations (\ref{eq:gw01}) \& (\ref{eq:gw_energy1})
and the inequality $(\rho_\phi/\rho_{\rm rad})_{\rm kin} \ggeq
(\rho_\phi/\rho_{\rm rad})_{\rm end}$,
\beq
\frac{\epsilon_g}{\epsilon_{\rm rad}} (\tau = \tau_{\rm eq})
\ggeq \frac{640}{3}
\left (\frac{V_{\rm HC}}{V_{\rm end}}\right )^3 g_p^{-1}.
\eeq
Since $V_{\rm HC}(\tau_{\rm eq})/V_{\rm end} > {\rm few}$, we obtain the conservative
bound
\beq
0.2 \ggeq \frac{\epsilon_g}{\epsilon_{\rm rad}} (\tau = \tau_{\rm eq})
> 2 \times 10^4 g_p^{-1}.
\eeq
The requirement that gravity waves are subdominant during nucleosynthesis
therefore translates into the following model independent constraint
on the number of particle species $g_p > 10^5$. We therefore conclude that
brane-inspired quintessential inflation 
(with inflationary particle production)
can only be accomodated if the number of particle
species through which the universe reheats is very large.

This situation can be remedied either by having a shorter kinetic
regime or if the inflaton were to have a post-inflationary equation
of state which is softer than $P_\phi = \rho_\phi$. Both these
features are accommodated by the potential
\beq
V(\phi) = V_0(\cosh{\tilde\alpha\phi/M_P} - 1),
\label{eq:pot1_again}
\eeq
earlier discussed in section \ref{sec:cosh}.
While brane-driven inflation proceeds along lines identical to
those examined earlier for exponential inflation, the post-inflationary
evolution in this model is radically different since reheating
takes place not due to gravitational particle production 
(as in the case of the exponential potential) but via scalar field
oscillations which commence when
$m^2 \equiv V'' = 8\pi V_0\tilde\alpha^2/M_4^2 = H^2$.
Prior to this epoch,
the expansion rate of the universe after the end of
inflation and before reheating, has the form
$a \propto t^{2/3(1+w_\phi)}$
where
\ber
w_\phi &=& \frac{\tilde\alpha^2}{3} - 1, ~~ \tilde\alpha \lleq \sqrt{6},\nonumber\\
w_\phi &=& 1, ~~ ~~~~~~~~\tilde\alpha \ggeq \sqrt{6}.
\eer
Restricting ourselves for the moment
to the stiff equation of state $w_\phi = 1$
we find for the energy density
\beq
\epsilon_{\rm g} = \frac{32}{3\pi} h_{\rm GW}^2 \epsilon_{\rm rad}
\left (\frac{H_{\rm kin}}{H_{\rm rh}} \right )^{2/3}
\eeq
where $H_{\rm kin}$ marks the commencement of the kinetic regime
and
$H_{\rm rh}^2 \simeq V''$
is the square of the Hubble parameter at the time of reheating.
The requirement that the gravity wave density satisfy nucleosynthesis
constraints leads to $\epsilon_{\rm g} \lleq 0.2 \epsilon_{\rm rad}$
and results in a firm lower bound on the value of $H_{\rm rh}$:
\beq
H_{\rm rh} \ggeq \frac{0.02}{\tilde\alpha}
(0.085 - \frac{0.69}{\tilde\alpha^2})\, {\rm GeV}
\label{eq:hubble1}
\eeq
where we have used the relationships (\ref{eq:fit}),
(\ref{eq:tension1}) \& (\ref{eq:tension2})
to establish (\ref{eq:hubble1}).
The temperature at reheating may be estimated from
\beq
T_{\rm rh}^2 \simeq \sqrt{\frac{3}{8\pi}}\zeta  M_4 H_{\rm rh}
\label{eq:hubble5}
\eeq
which results in the lower bound
\beq
T_{\rm rh} = (3V_0\tilde\alpha^2)^{1/4}
\ggeq \frac{3 \times 10^8}{\tilde\alpha^{1/2}}
\left (0.085 - \frac{0.69}{\tilde\alpha^2}\right )^{1/2} {\rm GeV},
\eeq
where $\zeta < 1$ is an indicator of the `efficiency' of reheating and
characterizes the fraction of the inflaton density converted into radiation
when the universe reheats, we assume $\zeta \simeq 1$ for simplicity.

From (\ref{eq:bog8}) - (\ref{eq:bog10}), (\ref{eq:hubble}) \& 
(\ref{eq:spectrum})
we find that the spectral energy density of the relic gravity wave
background is determined by the following set of equations (see Fig. 5)
\ber
\Omega_{\rm g}^{(\rm MD)}(\lambda) &=& \frac{3}{8\pi^3} h_{\rm GW}^2
\Omega_{0m}\left (\frac{\lambda}{\lambda_{h}}\right )^2,
~\lambda_{MD} < \lambda \leq \lambda_h\\
\Omega_{\rm g}^{(\rm RD)}(\lambda) &=& \frac{1}{6\pi}h_{\rm GW}^2\Omega_{0r},
~\lambda_{RD} < \lambda \leq \lambda_{MD}\\
\Omega_{\rm g}^{(\rm kin)} (\lambda) &=& \Omega_{\rm g}^{(\rm RD)}
\left (\frac{\lambda_{RD}}{\lambda}\right ),
~\lambda_{\rm kin} < \lambda \leq \lambda_{RD}
\eer
the superscripts `kin', `RD', `MD' refer to the
epoch when gravity waves in the given wavelength band were created,
the wavelength bands themselves are determined by
\ber
\lambda_h &=& 2cH_0^{-1} \simeq 1.8 \times 10^{28}h^{-1}{\rm cm},\\
\lambda_{MD} &=& \frac{2\pi}{3} \lambda_h\left
(\frac{\Omega_{0r}}{\Omega_{0m}}\right )^{1/2},\\
\lambda_{RD} &=& 4 ~\lambda_h
\left (\frac{\Omega_{0r}}{\Omega_{0m}}\right )^{1/2}\frac{T_{\rm MD}}
{T_{\rm rh}} = \frac{3.6\times 10^{18}}{T_{\rm rh} ({\rm GeV})}{\rm cm},\\
\lambda_{\rm kin} &=& c H_{\rm kin}^{-1}\left (\frac{T_{\rm rh}}{T_0}\right )
\left (\frac{H_{\rm kin}}{H_{\rm rh}}\right )^{1/3}  =
\left (\frac{\tilde\alpha}
{0.085 - 0.69/\tilde\alpha^2}\right )^{2/3}
0.004 \, T_{\rm rh}^{1/3} ({\rm GeV}) ~{\rm cm}.,
\label{eq:horizons}
\eer
The length scales $\lambda_{MD}$, $\lambda_{RD}$ and $\lambda_{\rm kin}$ are related
to the comoving Hubble radius at the start
of the matter-dominated, radiative and kinetic regimes respectively.
(We have used the relationship (\ref{eq:fit}) in establishing
$\lambda_{\rm kin}$.
We have also assumed $\Omega_{0m} \simeq 1$, $\Omega_{0r} \simeq 2.48 \times
10^{-5}$h$^{-2}$,
an analysis which includes the presence of a late-time accelerating
stage will be discussed in a companion paper.)

To summarize, we find that each of the three post-inflationary expansion
epochs leaves behind a distinct imprint
on the relic gravity wave background. As a result $\Omega_g(\lambda)$
(i) {\em decreases with wavenumber}
for modes created during matter domination,
(ii) {\em remains constant} for modes created during radiation
domination and (iii) {\em increases with wavenumber} for
modes created
during the kinetic regime.
The increase in amplitude of the gravity wave spectrum for wavelengths
shorter than $\lambda_{\rm RD}$ cm is expected to be a
generic feature
of inflationary models in which brane-induced damping is solely responsible
for driving inflation, since these models
enter the kinetic regime once the inflaton density has dropped below
the brane tension value $\lambda_b$.
In models without steep potentials such as braneworld models of
chaotic inflation, stage (iii) will be absent,
and the gravity wave spectrum will remain flat during the entire radiative
stage which will now extend from $\lambda \simeq \lambda_{\rm MD}$
until the shortest wavelengths.

We can also evaluate the total relic gravity wave
energy density which is given by
\ber
\epsilon_{\rm g} &=& \epsilon_{\rm g}^{({\rm kin})} + \epsilon_{\rm g}^{({\rm RD})}
+ \epsilon_{\rm g}^{({\rm MD})}, ~~~{\rm where}
\label{eq:gravsum}\\
\epsilon_g^{({\rm kin})} &=&
\frac{1}{\pi^2 a^4}\int_{k_{\rm RD}}^{2\pi/\tau_{\rm kin}} dk k^3
\vert\beta_{\rm kin}\vert^2,\\
\epsilon_g^{({\rm RD})} &=& \frac{1}{\pi^2 a^4}\int_{k_{\rm MD}}^{k_{\rm RD}}
dk k^3 \vert\beta_{\rm RD}\vert^2,\\
\epsilon_g^{({\rm MD})} &=& \frac{1}{\pi^2 a^4}\int_{2\pi/\tau}^{k_{\rm MD}}
dk k^3 \vert\beta_{\rm MD}\vert^2,
\label{eq:gravsum1}
\eer
substituting from (\ref{eq:bog8}) -- (\ref{eq:bog10}) we obtain
\ber
\epsilon_{\rm g}^{({\rm kin})} &=& \frac{32}{3\pi} h_{\rm GW}^2 \epsilon_{\rm rad}
\left (\frac{H_{\rm kin}}{H_{\rm rh}} \right )^{2/3}, \\
\epsilon_{\rm g}^{({\rm RD})} &=& \frac{2}{3\pi} h_{\rm GW}^2\epsilon_{\rm rad}
\log{\left (\frac{\pi\tau_{\rm MD}}{6\tau_{\rm RD}}\right )},\\
\epsilon_{\rm g}^{({\rm MD})} &=& \frac{1}{8\pi^3}h_{\rm GW}^2\epsilon_{\rm m}.
\eer
The dominant contribution to the gravity wave energy density clearly comes
from the first term in (\ref{eq:gravsum})
so that $\epsilon_{\rm g} \simeq \epsilon_{\rm g}^{({\rm kin})}$.
The nucleosynthesis bound
$\epsilon_{\rm g}^{({\rm kin})} \lleq 0.2 \epsilon_{\rm rad}$ can now be used
to get the upper limit $\Omega_{\rm g} \lleq 10^{-5}$.
We therefore find that the net gravity wave background
in brane-world inflationary models can be
significantly larger than that predicted in conventional
inflationary models ($\Omega_{\rm g} \sim 10^{-12}$) although it
is probably still too small to have significant cosmological implications.

\begin{figure}[h]
\centering
\resizebox{!}{5in}{\includegraphics{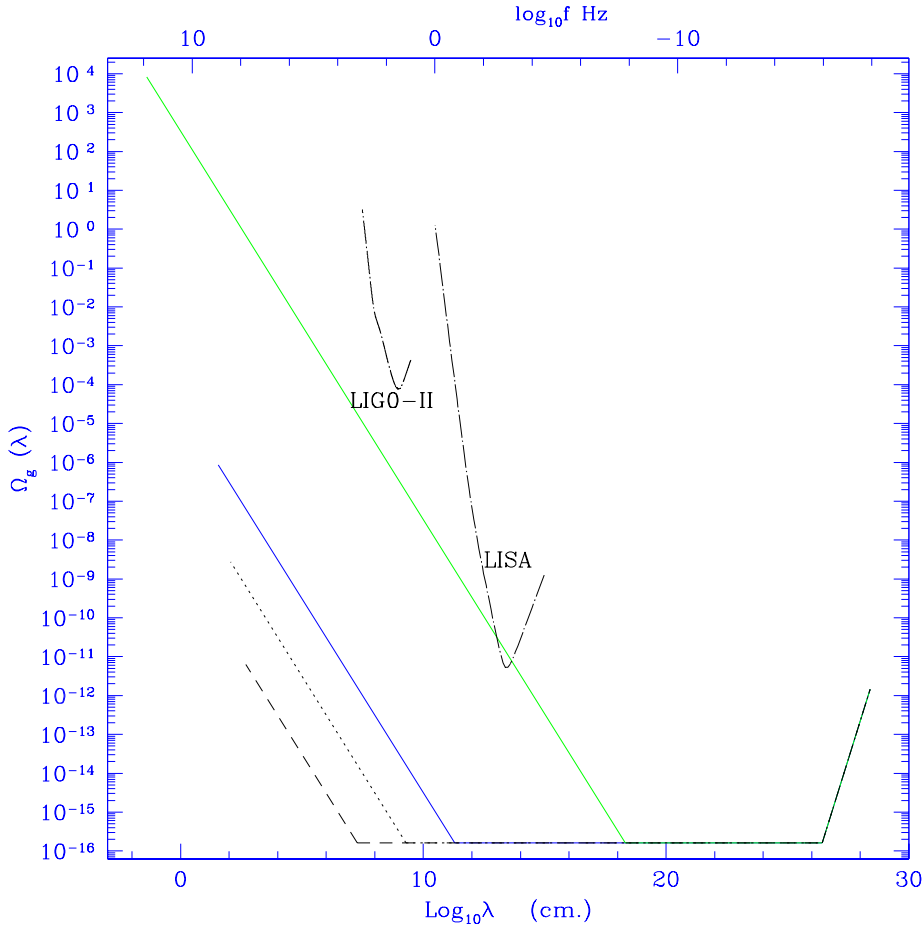}}
\caption{The COBE-normalized gravity wave spectrum is shown
for the cosine hyperbolic potential with 
$\lbrace\lambda_{\rm
RD},\lambda_{\rm kin}\rbrace = \lbrace 2\times 10^{11} {\rm cm},
36 {\rm cm}\rbrace, \lbrace 2\times 10^{9}{\rm cm}, 114{\rm cm}
\rbrace$ and $\lbrace 2\times 10^{7} {\rm cm}, 488 {\rm
cm}\rbrace $ these values correspond to reheating temperatures:
$T_{\rm rh} \simeq 2\times 10^7$ GeV, $2\times 10^9$ GeV,
$2\times 10^{11}$ GeV and $\tilde\alpha = 16, 8, 4$ respectively.  The dark
solid line, dotted, and short dashed line correspond to the three
cases with $\tilde\alpha=16, 8~ \& ~4$, respectively.
For comparison we also show
the gravity wave background for the exponential potential
(gray solid line).
In this model radiation is created
quantum mechanically during inflation, resulting in a low
temperature at the commencement of the radiative stage $T_{\rm eq}
\simeq 0.3 g_p^{1/2} $ GeV for $\tilde\alpha = 16$ (see Sec. ~\ref{sec:exp}). 
The corresponding value of the (comoving) Hubble radius at
the commencement of the radiative epoch is $\lambda_{\rm RD} \simeq
 2\times 10^{18}$ cm assuming $g_p \simeq 50$ and the (comoving) Hubble radius
 at the commencement of the
kinetic epoch is $\lambda_{\rm kin} = cH_{\rm kin}^{-1}\left (
\frac{T_{\rm kin}}{T_0}\right )
\simeq 0.04$ cm. The large gravity wave energy density predicted by this model
appears to be in serious conflict with primordial nucleosynthesis constraints.
The GW spectra assume the present value of the Hubble
constant $H_0 = 70 {\rm km/s/Mpc}$.
The chained lines marked LISA and LIGO-II are the expected
sensitivity curves of the proposed Laser Interferometer Space
Antenna and second phase of the Laser Interferometric gravity wave
observatory.  }
\end{figure}

\section{Discussion/Conclusions}

An important feature of braneworld models based on the Randall-Sundram
ansatz is the increased rate of expansion of the universe which assists
slow roll and thereby considerably enlarges the family of potentials
which can contribute to inflation.

In the present paper we have made a detailed
analysis of a family of steep inflationary potentials which includes:
(i) $V(\phi) \propto  e^{\tilde\alpha\phi/M_P}$,
(ii) $V(\phi) \propto [\cosh{(\tilde\alpha\phi/M_P)} - 1]^p$,
(iii) $V(\phi) \propto \phi^{-\tilde\alpha}$.
In all three cases we
find that the universe succesfully inflates.
In the case of (ii) \& (iii) the scalar field, in a limited region of parameter
space, can play the dual role of being both the inflaton and dark energy.
A generic feature of inflationary models with steep potentials is that
the post-inflationary epoch is characterised by a
prolonged kinetic regime during which the effective equation of state
is $w_\phi \simeq 1$ and the universe expands as $a(t) \propto t^{1/3}$.
This regime lasts until 
the energy density of
radiation (created quantum mechanically during inflation)
becomes equal to the matter density.

For models (ii) and (iii)
a long duration kinetic regime leads to the possibility that inflation may
resume prior to reheating ! Viable models in which 
this does not occur and in which the universe begins to accelerate 
close to the present cosmological epoch (neither much earlier nor much later)
can also be constructed, but for a very limited region of parameter space.
In these models {\em both} inflation and
dark energy are generated by the very same scalar field. 

Inflationary models leave behind a distinct imprint on the
stochastic gravity wave background.
Since the gravity wave spectrum is acutely sensitive to
the post-inflationary
equation of state of matter, a long-duration kinetic regime
results in a gravity wave spectral density which
increases with wavenumber $\tilde \epsilon_{\rm g} \propto k$
for wavelengths shorter than the comoving Hubble radius at the commencement of
the radiative regime.
For models (i) -- (iii), with reheating generated {\em solely} on account of
inflationary particle production,
the kinetic regime is excessively long and the resulting 
`blue tilted' energy density
of gravity waves can exceed the
radiation density violating cosmological nucleosynthesis
considerations. In models allowing conventional reheating such as
(ii), the duration of the kinetic regime is related to parameters in
the potential. Such models
leave behind a distinct signature on the relic
gravity wave background without necessarily violating
nucleosynthesis constraints.
The only potential capable of being both the inflaton and quintessence
without generating an unacceptably large gravity wave background
is $V(\phi) = V_0\cosh{\tilde\alpha\phi/M_P}$, provided the
parameters 
$V_0\tilde\alpha^2 \simeq H_{\rm rh}^2$ satisfy the 
nucleosynthesis constraints
(\ref{eq:hubble1}). However this model involves considerable 
fine tuning since the value of the parameter $V_0$ must be set equal to
the current value of the cosmological constant 
$V_0 \simeq 10^{-47} {\rm GeV}^4$.

To summarize, we have shown that models of `quintessential inflation' 
are possible to construct within the context of braneworld cosmology.
However, most models of quintessential inflation also
generate a relic gravity wave background which can be several orders of
magnitude larger than in conventional models.
In the case of inflationary models with steep potentials,
the relic gravity wave background is an extremely potent probe
which can be used both to rule out models
as well to (indirectly) confirm the reality of extra dimensions.

\[ \]

{\bf Acknowledgments:}\\

We thank Bruce Allen, Massimo Giovannini, Greg Huey, 
Andrew Liddle, James Lidsey, Tarun Deep Saini, B.S. Sathyaprakash,
Alexei Starobinsky and
Tabish Qureshi for useful discussions.\\
Helpful comments from an anonymous referee are also acknowledged.

\end{document}